\documentclass[12pt]{article}

\usepackage{amsmath,amssymb}
\usepackage{graphics}
\numberwithin{equation}{section}

\newcommand{\alp}{\alpha}
\newcommand{\bt}{\beta}

\newcommand{\dlt}{\delta}

\newcommand{\ep}{\epsilon}
\newcommand{\vep}{\varepsilon}
\newcommand{\tht}{\theta}

\newcommand{\kp}{\kappa}
\newcommand{\lmd}{\lambda}
\newcommand{\Lmd}{\Lambda}
\newcommand{\sgm}{\sigma}
\newcommand{\Sgm}{\Sigma}
\newcommand{\vph}{\varphi}
\newcommand{\omg}{\omega}

\newcommand{\Lg}{\Lambda_{\rm G}}
\newcommand{\Lc}{\Lambda_*}

\def\ignore#1{{}}

\newcommand{\be}{\begin{equation}}
\newcommand{\ee}{\end{equation}}
\newcommand{\bea}{\begin{eqnarray}}
\newcommand{\eea}{\end{eqnarray}}
\newcommand{\eql}{\!\!\!&=\!\!\!&}

\newcommand{\defa}{\!\!\!&\equiv\!\!\!&}

\newcommand{\bpx}{\begin{pmatrix}}
\newcommand{\epx}{\end{pmatrix}}

\newcommand{\tl}[1]{\tilde{#1}}
\newcommand{\bdm}[1]{{\mbox{\boldmath $#1$}}}
\newcommand{\tr}{{\rm tr}}

\newcommand{\der}{\partial}

\newcommand{\hc}{{\rm h.c.}}
\newcommand{\ie}{{\it i.e.}}

\newcommand{\vev}[1]{\langle #1 \rangle}

\newcommand{\brkt}[1]{\left( #1 \right)}
\newcommand{\brc}[1]{\left\{ #1 \right\}}
\newcommand{\sbk}[1]{\left[ #1 \right]}
\newcommand{\abs}[1]{\left| #1 \right|}

\renewcommand{\Re}{{\rm Re}\,}
\renewcommand{\Im}{{\rm Im}\,}

\newcommand{\cL}{{\cal L}}
\newcommand{\cM}{{\cal M}}
\newcommand{\cN}{{\cal N}}
\newcommand{\cO}{{\cal O}}
\newcommand{\cP}{{\cal P}}

\newcommand{\cV}{{\cal V}}
\newcommand{\cW}{{\cal W}}

\begin{document}
\baselineskip=16pt
\begin{titlepage}
\begin{flushright}
{\small OU-HET 594/2007}\\
\end{flushright}
\vspace*{1.2cm}

\begin{center}

{\Huge\bf iGUT}\\
\vspace{5mm}
{\Large\bf 
$-$Grand Unified Theory on Interval$-$ } 

\lineskip .75em
\vskip 1.5cm

\normalsize
{\large Naoyuki Haba},
{\large Yutaka Sakamura},
and
{\large Toshifumi Yamashita}

\vspace{1cm}

{\it Department of Physics, 
 Osaka University, Toyonaka, Osaka 560-0043, 
 Japan} \\

\vspace*{10mm}

{\bf Abstract}\\[5mm]
{\parbox{13cm}{\hspace{5mm}
%

We investigate a construction of 
 five-dimensional (5D) grand unified theories (GUTs)
 on an interval, which we call iGUTs. 
We analyze supersymmetric $SO(10)$ iGUT 
 as an example, where 
 the gauge multiplet is 
 spread over the 5D bulk. 
The $SO(10)$ is directly reduced to 
 the standard model gauge symmetry
 through the interval boundary conditions.
Notice that this rank reduction is impossible
 in case of GUTs on orbifolds. 
 Four scenarios are possible 
 according to 
 locations (bulk or brane) of Higgs 
 and matter
 fields. 
We investigate the gauge-coupling unification, 
 the proton decay, the $SO(10)$ GUT features 
 such as $t$-$b$-$\tau$ unification and so on 
 in each scenario. 
We also comment on the flavor phenomenology. 
}}

\end{center}

\end{titlepage}

\section{Introduction}

A supersymmetric (SUSY) grand unified theory
 (GUT) is an attractive candidate as an underlying 
 theory of the standard model (SM).
The strongest reason is that the three SM gauge couplings 
 seem unified at a high energy 
 scale, $\Lg\simeq 2 \times 10^{16}$ GeV,  
 which is the so-called GUT scale. 
However, recent precise measurements
 of the QCD gauge coupling show a small but
 finite deviation from the predicted value
 of the unification\cite{PDG}. 
Also, some theoretical problems exist
 in the four-dimensional (4D) minimal SUSY $SU(5)$ GUT.
For example,
 the triplet-doublet splitting in the Higgs multiplets 
 should be realized by an unnatural fine-tuning
 of $\cO{(10^{14})}$. 
So 
 people pay attention to 
 five-dimensional (5D) $SU(5)$ GUT on
 an orbifold\cite{Kawamura,Hall},  
 which 
 realizes the 
 gauge symmetry breaking and 
 the triplet-doublet splitting simultaneously
 through boundary conditions (BCs) of
 the orbifold. 
We do not need to introduce 
 adjoint Higgs fields
 to break the GUT gauge symmetry which 
 usually violate the $R$-symmetry explicitly
 in the superpotential. 
Furthermore, 
 a precise gauge coupling unification (GCU)
 can be realized by taking the compactification scale 
 lower than the GUT scale~$\Lg$\cite{Hall:2001xb}.
This situation corresponds to 
 take the triplet Higgs masses lighter than $\Lg$ in the
 4D GUTs, 
 which however causes too 
 rapid proton decay\cite{Murayama:2001ur}. 
This problem is avoidable 
 in the 5D setup, 
 since the triplet Higgs fields get heavy masses
 with their chiral partners without 
 violating the $R$-symmetry\cite{Hall}. 
The $R$-symmetry is valid to forbid 
 problematic dimension-five operators in general.  
Thus
  the 5D GUTs on the orbifold are attractive  from   
  these phenomenological points of view.                                 
However 
 the rank of the GUT gauge symmetry cannot 
 be reduced on the orbifold 
 BCs.\footnote{Precisely speaking, 
 a rank reduction can be possible in 
 a gauge-Higgs unification scenario\cite{GHU} 
 which is not considered in this paper.}
So the GUTs with higher ranks than the SM
 must have extra remaining gauge
 symmetries, which should be broken by
 introducing extra elementary Higgs fields.
Thus, if we would like to consider 
 $SO(10)$ GUT, which 
 unifies quarks and leptons in a single
 multiplet, 
 we must introduce 
 additional GUT-symmetry breaking Higgs fields. 
For example, 
 in Ref.\cite{Dermisek:2001hp}, 
 the orbifold BCs break 
 $SO(10)$ into the Pati-Salam gauge group, 
 which is subsequently broken to the SM 
 by vacuum expectation values (VEVs) 
 of additional Higgs fields. 
Another setup 
 is a six-dimensional spacetime 
 where orbifold BCs break $SO(10)$ 
 to the SM gauge group times
 an extra $U(1)$ which must be broken by  
 additional Higgs fields again\cite{SO(10)}. 
Anyhow, 
 the existence of extra gauge groups is 
 inevitable in the orbifold GUTs.

Recently,
 some people consider an interval
 instead of the orbifold for the 
 compactification space in 5D models\cite{Csaki:2003dt}. 
It provides larger class of BCs than the orbifold,
 which are consistent with  
 the action principle. 
The tree-level unitarity is also maintained for certain
 interval BCs\cite{Csaki:2003dt,Uekusa},\footnote{The unitarity
 under the orbifold BCs in the flat extra dimension 
 is guaranteed
 by an equivalence theorem\cite{Abe:2004wv}.}  
 part of which 
 can be obtained from the orbifold by introducing non-dynamical 
 Higgs fields (which we call 
 {\it fake} Higgs fields)
 on the orbifold boundaries and taking
 their VEVs to infinity\cite{Csaki:2003dt,Nomura:2001mf}.
It is remarkable that 
 the rank of the gauge group is reducible 
 by the interval BCs in contrast 
 to the orbifold. 
We stress that 
 the interval can take BCs which the orbifold
 cannot realize. 
For this reason,
 the interval is useful for the extra-dimensional 
 model building 
 in various contexts.
However, most of the works on the interval use the interval BCs 
in models of the electroweak symmetry breaking, namely, 
the Higgsless models\cite{Csaki:2003dt,Csaki:2003zu} 
 or the gauge-Higgs unification models\cite{Agashe:2004rs}. 
The application of the interval BCs to the GUT-symmetry breaking 
has not been studied so far, 
except for the trinification model\cite{Carone:2004rp}.

In this paper, 
 we investigate a construction of 5D ${\cal N}=1$ SUSY GUTs on
 the interval, which we call iGUTs. 
The gauge multiplets are set to be spread 
 over the 5D bulk. 
The rank of the GUT gauge symmetry is reduced 
 through the interval BCs 
 differently from the orbifold. 
In Section~\ref{Sec:BCbyHand}, 
 we consider $SO(10)$ iGUT, 
 and discuss four  scenarios depending on
 locations of the Higgs and matter fields in the extra
 dimension (bulk or boundary). 
The discussion on the GCU in the orbifold GUTs\cite{Hall:2001xb} 
 is applied for these scenarios in Section~\ref{Sec:GCU}.
In Section~\ref{Sec:BC},
 we review a construction of 
 interval BCs by introducing the {\it fake} Higgs
 fields on the boundaries and taking 
 their VEVs to infinity. 
Useful formulae are collected in Appendices. 
Section~\ref{summary} is devoted to the
 summary and discussions.

\section{$\bdm{SO(10)}$ iGUT}
\label{Sec:BCbyHand}

Let us consider 
 the 
 $SO(10)$ iGUT 
 with the flat metric.      
In this section,
 we 
 impose interval BCs by hand 
 at the two end points, $y=0$ and $\pi R$,
 which break $SO(10)$ to the SM gauge symmetry.
Here $y$ is the 5th dimensional coordinate, and 
 we call these two end points as branes or boundaries 
 in the following discussions. 
We should remind that       
 any orbifold BCs cannot realize the direct 
 GUT-symmetry
 breaking of $SO(10)\rightarrow$ SM. 
The minimum field content is 
 the gauge multiplet~${\bf 45_G}$, 
 matter multiplets~${\bf 16_M}$, and 
 a Higgs multiplet~${\bf 10_H}$. 
The doublet Higgs fields of the
 minimal SUSY SM (MSSM) are contained 
 in ${\bf 10_H}$. 
There are no GUT-symmetry breaking
 Higgs in this field content. 
The gauge multiplet is spread 
 over the 5D bulk, and
 the matter and Higgs fields are
 either bulk or brane fields. 
Realization of the following 
 BCs by use of the {\it fake} Higgs fields 
 will be discussed in Section~\ref{Sec:BC}.

As for the gauge multiplet, 
 we take the Neumann (Dirichlet) BCs for the SM ($SO(10)$/SM) gauge
 fields $A_\mu^a$ ($A_\mu^{\hat a}$) on the $y=\pi R$ boundary, 
 where $a$ ($\hat{a}$) denotes the SM
 ($SO(10)$/SM) gauge index. 
Thus, the gauge symmetry is reduced to the SM one 
 at $y=\pi R$. 
On the other hand, we take the Neumann BCs for 
 all components of the $SO(10)$ gauge multiplet at $y=0$.
Therefore the BCs at both the branes are given as 
\begin{equation}
 \partial_y A_\mu^a = \der_y A_\mu^{\hat a} = 0,  \;\;\; (y=0), \;\;\;\;\;\; 
 \partial_y A_\mu^a = A_\mu^{\hat{a}} = 0,  \;\;\; (y=\pi R).                 
\label{gaugeBC}
\end{equation}
The BC at $y=0$ makes 
 all 
 components of $A_y$ 
 heavy. 
Thus, 
 there are no physical degrees of freedom in $A_y$, 
 which are absorbed into
 the longitudinal components    
 of massive gauge fields.  
So we focus on $A_\mu$ in the following discussions.  
The BCs in Eq.(\ref{gaugeBC}) 
 give the lightest mode of $A_\mu^{\hat a}$ a mass of $1/(2R)$, 
 while that of $A_\mu^{a}$ remains massless. 
It means that the 4D effective theory 
 has the SM gauge symmetry. 
This is a kind of the Higgsless
 breaking of the GUT symmetry.

We should determine 
 the locations of the matter and Higgs fields 
 for the discussion of phenomenological
 issues, such as the triplet-doublet splitting,
 the proton decay, the GCU, and so on.
There are the following four scenarios according to 
 the 5D locations of
 the MSSM Higgs doublets 
 and matter fields.

\subsection{Brane Higgs and brane matter}
\label{Sec:brane-brane}

The first scenario is 
 putting both the Higgs doublets and matter
 fields on the $y=\pi R$ brane. 
The gauge symmetry on this brane is already reduced 
 to the SM one, so that 
 $SO(10)$-incomplete multiplets
 and $SO(10)$-breaking interactions can be
 introduced on it. 
Some features of the $SO(10)$ GUT
 are lost in this setup, 
 for example, 
 the $t$-$b$-$\tau$ unification and  
 unification of the right-handed neutrinos and other matters. 
And 
 the charge quantization $Q(p^+)=-Q(e^-)$ 
 nor the automatic anomaly cancellation 
 of $SO(10)$ are not guaranteed.  
This setup 
 seems not so attractive, 
 however, 
 has the following 
 good features. 
Absence of 
 triplet Higgs fields makes 
 the dangerous dimension-five proton
 decay operators mediated by
 them vanish. 
Dimension-six operators are also absent 
 since 
 the coset space gauge fields $A_\mu^{\hat a}$ do not 
 couple to the brane matter fields due to 
 no overlap at $y=\pi R$ brane.\footnote{
The extra-dimensional components 
 $A_y^{\hat a}$ do not appear in the 4D minimal coupling
 with the brane fields. 
Although higher-derivative interactions of them which are 
 localized on
 the branes might be possible\cite{Nomura:2001mf, Hebecker:2002rc}, 
 we assume their absence in this paper,
 for simplicity.  
} 
As for 
 intrinsic dimension-five operators
 suppressed by the cutoff scale,
 they are (almost) forbidden by imposing 
 the (approximate) $R$-symmetry. 
Remind that 
 this is impossible in the 4D setup, since 
 the $R$-symmetry is broken at the
 GUT scale through the triplet and adjoint Higgs masses.

The $R$-symmetry is set to be 
 broken only in the hidden (SUSY-breaking) sector. 
There are the following three options  for
 the location of the 
 hidden sector. \\

\vspace{-4mm}
\noindent
{\bf Hidden sector localized on the $\bdm{y=0}$ brane:}
The SUSY flavor problem can be solved 
 by the gaugino 
 mediation\cite{Kaplan:1999ac}. 
Recalling that the gravitino mass is 
 $m_{3/2}\simeq F/M_P$ 
 ($M_P\simeq1.2\times10^{19}{\rm GeV}$: 4D Planck scale, 
  $F$: order parameter of SUSY breaking), 
 the gaugino mass is expressed as  
 $M_{1/2}=F/(2\pi R\Lc^2) \simeq m_{3/2}\times(\dlt^2/\epsilon)$.  
Here $\ep\equiv \Lc/M_P$ ($\Lc$: 5D cutoff scale), and 
 $\delta \equiv 1/\sqrt{2\pi R\Lc}$ is the volume suppression factor, 
 which must be less than one 
 if the 5D description is valid.\footnote{
We assume that brane-localized couplings normalized by $\Lc$ 
 are of $\cO(1)$, while the bulk gauge coupling constant is 
 somewhat large in order to realize the suitable value of 
 the 4D gauge coupling constants. 
} 
Since $\Lc$ is at most the 5D Planck scale~$M_5$ that is 
 related to $M_P$ through $M_P^2=2\pi RM_5^3$, 
 these quantities 
 satisfy the following relation. 
\be
  0< \epsilon \leq \delta < 1, 
\ee
where the equality holds when $\Lc=M_5$. 
The other soft SUSY breaking masses 
 are induced from 
 the gaugino mass 
 through
 the renormalization group equations~(RGEs)  
 though they are small at the compactification scale, 
 and then the SUSY flavor problem is 
  solved\cite{Kaplan:1999ac}. 
Therefore the soft SUSY masses are of the order of 
 the gaugino mass in the low energy.\footnote{
The anomaly mediation\cite{Randall:1998uk}
 is effective in the case of 
 heavy gravitino mass with 
 $\delta^2/\epsilon \leq 10^{-2}$.
} 
To be more concrete, in the leading-log approximation, 
 flavor independent soft squared masses are generated 
 through the gaugino loop as 
\begin{equation}
 \tilde m^2
 =8Tg_4^2M_{1/2}^2 \frac{\ln(M_c/M_{\rm SUSY})}{16\pi^2}
 \sim M_{1/2}^2,
\label{GauginoCont}
\end{equation}
 where $T$ is a group factor being of order 1, $g_4$ is the 
 4D effective gauge coupling and $M_c\equiv 1/R$ is the compactification 
 scale.

In this scenario, the 
 $\mu$-term  is difficult to
 be induced from the hidden sector. 
The simplest example of generating $\mu$ is to introduce       
 a gauge singlet field on the $y=\pi R$ brane whose VEV 
 becomes the $\mu$-term\cite{NMSSM}. \\ 

\vspace{-4mm}
\noindent
{\bf Hidden sector localized on the $\bdm{y=\pi R}$ brane:}
The SUSY flavor problem is revived again as in the 4D GUTs. 
Thus 
 another flavor-independent SUSY mediation 
 must be introduced and 
 dominate the gravity mediation  
 for the suitable soft SUSY breaking masses.

The $\mu$-term 
 can be induced by 
 a direct coupling 
 between the hidden sector's spurion field $X$ 
  and the Higgs fields as
 $X^\dagger H_uH_d$ in the K\"{a}hler
 potential\cite{Hall11}. 
This case tends to realize a large 
 $\mu \sim m_{3/2}\times\ep^{-1}$ so that 
 the coupling of $X^\dagger H_uH_d$ should be 
 tuned to be small in order for $\mu$ to be
 the same order as
 $M_{1/2}\sim m_{3/2}\times(\dlt^2/\epsilon)$.\footnote{
In a similar way, the coupling of $X^\dagger XH_uH_d$ 
 should be tuned to be small 
 to avoid a large $B$-parameter\cite{Kaplan:1999ac}.
} 


\vspace{2mm}
\noindent
{\bf Radion $\bdm{F}$-term:}
The SUSY breaking can be induced  
 through 
 the radion $F$-term\cite{MartiPomarol,KaplanWeiner}, which  
 is equivalent to the Scherk-Schwarz 
 SUSY breaking\cite{SS}
 in the flat metric\cite{MartiPomarol,AS2}.\footnote{
These are also equivalent to putting constant superpotentials 
 in the branes\cite{AS2,BFZ}. 
} 
The gaugino masses are induced from the radion $F$-term,
 which derives all soft SUSY masses through
 the RGEs as above.  
The gravitino mass is the same order as 
 the gaugino mass in this setup 
 as $M_{1/2}\sim m_{3/2}$. 

The $\mu$-term might be obtained 
 by introducing an extra singlet.\footnote{
There is no direct interaction $T^\dagger H_uH_d$ 
 ($T$: radion) on the branes in the flat metric. 
}  

\subsection{Brane Higgs and bulk matter}
\label{Sec:brane-bulk}

The second scenario is 
 putting the matter fields in the bulk while the doublet 
 Higgs fields remaining 
 on the $y=\pi R$ brane. 
As in the first option, 
 the anomaly cancellation
 of $SO(10)$   
 is not automatic. 
We denote a matter hypermultiplet ${\bf 16_M}$ 
 as $({\bf 16}, {\bf 16^c})$, 
 where ${\bf 16}$ and ${\bf 16^c}$ correspond 
 to $\cN=1$ SUSY chiral multiplets. 
We take the BCs as 
\begin{equation}
 \partial_y {\bf 16} = {\bf 16^c} = 0                                          
\end{equation}
at both boundaries. 
It is worthwhile to notice that the BCs are compatible 
 with the $SO(10)$ bulk gauge symmetry, 
 in contrast to the orbifold BCs.%
\footnote{
For instance, in the $SU(5)$ orbifold model where the BCs 
 breaks $SU(5)$ to the SM symmetry, 
 $D^c$ and $L$ in the ${\bf\bar5}$ multiplet must have  
 opposite parities. Thus, 
 it is impossible for both components to serve 
 zero-modes from a single ${\bf\bar5}$ bulk hypermultiplet, 
 but two multiplets should be introduced\cite{Hall}.
\label{IncompUnif}
}
Because the Higgs fields are localized on the $y=\pi R$ brane, 
 the Yukawa interactions have to be localized on the brane,
 allowing us to introduce appropriate couplings  
 of the MSSM.

Due to  
 the absence of the triplet Higgs fields,
 dimension-five 
 proton decay operators induced by
 them  are absent, and 
 the intrinsic 
 dimension-five 
 proton decay operators are suppressed 
 by imposing the (approximate) 
 $R$-symmetry  
 as in the scenario in Section 2.1. 
On the other hand, the dimension-six proton decay
 processes mediated by the heavy gauge bosons exist 
 because 
 the matter fields couple to $A_\mu^{\hat a}$ in the bulk.%
\footnote{
In the orbifold models, 
 such dimension-six operators are absent 
 because $D^c$ and $L$ ($Q$ and $(U^c,E^c)$) reside 
 in different ${\bf\bar5}$ (${\bf10}$) multiplets, 
 as mentioned in the footnote \ref{IncompUnif}.
}
The experimental lower bound on the lightest KK mass for $A_\mu^{\hat a}$,
 which is a half of the compactification scale, $1/(2R)$,  
 is estimated using a formula in Ref.\cite{Murayama:2001ur} as
\begin{equation}
 \frac{1}{2R} \geq
  3\times 10^{15}\ {\rm GeV}
  \left(\frac{g_1^{\hat a}}{g_4}\right)
  \left(\frac{\tau_p(p\rightarrow e\pi^0)}{1.6\times 10^{33}{\rm yrs}}
  \right)^{1/4}
  \left(\frac{\abs{\alpha_H}}{0.01({\rm GeV})^3}\right)^{1/2},
\label{minMX}
\end{equation}
where $g_4$ is the unified gauge coupling constant in the effective 
 4D theory, $\tau_p(p\rightarrow e\pi^0)$ is the lower bound on 
 the proton lifetime whose present value is 
 $1.6\times 10^{33}{\rm yrs}$\cite{PDG},\footnote{
 A more stringent bound, $5.3\times 10^{33}{\rm yrs}$, 
 has been reported in Ref.\cite{PDproceeding}.} and 
 $\alpha_H$ is a constant of a nucleon-to-vacuum matrix element
 which would be between $0.003$ and 
 $0.03$\cite{Aoki:2006ib}. 
It should be noticed that 
 the coupling of the $n$-th KK mode for $A_\mu^{\hat a}$, 
 $A_\mu^{\hat{a}(n)}$, 
 to the matter fields is a new parameter indicated as $g_n^{\hat a}$,   
 which is calculated as an overlap integral of wave functions 
 of the matter fields and $A_\mu^{\hat{a}(n)}$, and thus 
 depends on the localization of the matter fields.
The localization of a bulk matter field
 can be realized by a parity-odd bulk mass, $m$, 
 which makes the wave function of the zero mode 
 have an exponential profile, $\exp(m y)$. 
It is straightforward to calculate the overlap integral among two 
 wave functions of the matter fields 
 and $A_\mu^{\hat{a}(n)}$, or that of the zero mode of $A_\mu^a$. 
Then we obtain the ratio between the former and the latter as 
\begin{equation}
 \frac{g_n^{\hat a}}{g_4}=
 2\sqrt2
 \frac{mR\left\{-(-1)^n(2n-1)
 -4e^{-2\pi mR} mR\right\}(\coth(\pi mR)+1)}
      {(2n-1)^2+(4mR)^2}.
\label{gn}
\end{equation}
For instance, the ratio for the lightest mode~$A_\mu^{\hat{a}(1)}$ 
 is calculated  
 as $g_1^{\hat a}/g_4=\sqrt2$ for the $y=0$
 brane-localized matters ($m\to-\infty$), 
 $g_1^{\hat a}/g_4=2\sqrt2/\pi=0.90$ for the matter with the  
 flat profile ($m=0$), 
 and $g_1^{\hat a}/g_4=0$ for the $y=\pi R$ 
 brane-localized matters ($m\to\infty$).
In reality, 
 every $A_\mu^{\hat{a}(n)}$ also mediates the proton decay, 
 though its contribution is suppressed by $(2n-1)^{-2}$ 
 compared to that of $A_\mu^{\hat{a}(1)}$ due to the heavier mass. 
Summing up those contributions, 
 we find that the effective coupling is given as
\begin{equation}
 \left(\frac{g^{\hat a}_{\rm eff}}{g_4}\right)^2\equiv
 \sum_n \left(\frac{g_n^{\hat a}}{g_4}\right)^2\frac{(1/2)^2}{(n+1/2)^2}=
 \pi
 \frac{2(\cosh(2\pi mR)-1)-\sinh(2\pi mR) +2\pi mR e^{-2\pi mR}}
      {32mR\sinh(\pi mR)}.
\label{geff}
\end{equation}
Then, 
 we effectively have 
 $g_{\rm eff}^{\hat a}/g_4=\sqrt{3\zeta_R(2)/2}=1.57$ 
 for the $y=0$ brane-localized matter, 
$g_{\rm eff}^{\hat a}/g_4=\sqrt{15\zeta_R(4)/2\pi^2}=0.91$ for
 the matter with the flat profile, and  
 $g_{\rm eff}^{\hat a}/g_4=0$ 
 for the $y=\pi R$ brane-localized matter.    
Here $\zeta_R(x)$ is the Riemann's zeta function.

In this way,
 the value of 
 $g_{\rm eff}^{\hat a}/g_4$ becomes small when
 the 1st and 2nd generation wave functions
 are localized around  $y=\pi R$, and then 
 the proton decay is strongly suppressed,   
 while 
 the smallness of these generation  masses 
 should be realized 
 by small Yukawa couplings on the brane 
 (or by some mechanism, 
 for example,
 the Froggatt-Nielsen (FN) mechanism\cite{Froggatt}).  
On the other hand, 
 if we want to 
 reproduce the fermion mass hierarchy
 by  the bulk matter localizations\cite{ArkaniHamed:1999dc},   
 the 1st and 2nd generation matter fields 
 should be localized around the $y=0$ brane. 
In this case,
 the value of $g_{\rm eff}^{\hat a}$, 
 and thus the dimension-six proton decay,
 are enhanced.   
As will be shown in Section~\ref{Sec:GCU},
 the precise GCU might need unknown 
 extra fields in the brane Higgs scenarios, so  
 the compactification scale~$1/R$ 
 cannot be determined at the present stage. 
When this mass is of the order of the GUT scale,  
 the decay rate of the process $p\to e\pi$ is enhanced 
 by a factor $6$ compared to the minimal $SU(5)$ model.
Anyhow, we should 
 notice that the 
 bulk matter profiles cannot
 explain all fermion mass hierarchies and
 flavor mixings only by themselves 
 due to the bulk $SO(10)$-symmetry. 
%

The SUSY flavor problem is not solved 
 due to the existence of the bulk matter fields. 
Neither 
 the hidden sector on the $y=0$ brane nor
 $y=\pi R$ brane can solve it.
The radion $F$-term also induces 
 the SUSY flavor problem due to 
 the generation dependent 
 bulk matter profiles\cite{ACJO}. 
In some context it can be solved due to
 suitable localizations 
 of the matter fields,
 as analysed in Section~\ref{Sec:bulk-bulk}, but, in principle,  
 another flavor-independent SUSY mediation 
 must be introduced and 
 dominate the gravity mediation  
 for the  suitable soft SUSY breaking 
 parameters. 

As for the $\mu$-term, 
 the situation is the same as Section 2.1. 
The interaction $X^\dagger H_uH_d$ can induce
 the suitable value of $\mu$ 
 when the hidden sector is
 localized on the $y=\pi R$ brane. 

\subsection{Bulk Higgs and brane matter}

The third scenario is putting 
 the ${\bf 10_H}$ Higgs hypermultiplet in the
 bulk whereas the matter fields 
 on the $y=\pi R$ brane.\footnote{
A case of the matter fields 
 localized on 
 the $y=0$ brane cannot reproduce 
 the realistic fermion mass spectrum.
}
As the first option in Section 2.1, 
 the introduction of the matter fields on 
 the $SO(10)$-breaking brane means that 
 the charge quantization nor the automatic anomaly cancellation 
 are no longer 
 guaranteed.
Denoting the hypermultiplet ${\bf 10_H}$ as 
 $(H, H^c)$ ($H^c$: chiral partner), 
 we take BCs for ${\bf 10_H}$ as 
\begin{equation}
 \partial_y H = H^c = 0,  \;\;\; (y=0), \;\;\;\;\;\; 
 \partial_y H_D = H_T = H_D^c = \partial_y H_T^c = 0,  \;\;\; (y=\pi R),
\label{HiggsBC}
\end{equation}
where $H^{(c)}=(H_T^{(c)}, H_D^{(c)})$ 
 with $H_T^{(c)}$ ($H_D^{(c)}$) being the 
 triplet (doublet) Higgs field. 
Here we omit an index that labels 
 two different Higgs fields, {\it i.e.}
 one forms the up-type Yukawa interactions
 and the other does the down-type ones. 
The triplet-doublet splitting
 is realized through these BCs similarly
 to the 5D $SU(5)$ GUT
 on the orbifold\cite{Kawamura}.

Again, 
 although the $SO(10)$-relations such as 
 the $t$-$b$-$\tau$ unification are lost,  
 appropriate Yukawa interactions and Majorana masses
 of the right-handed neutrinos 
 can be introduced 
 on the $y=\pi R$ brane. 
Since the triplet Higgs fields, $H^T$,
 have the Dirichlet BC,
 they do not couple with the brane-localized 
 quarks and leptons. 
The $R$-symmetry forbids 
 the dangerous intrinsic dimension-five proton decay
 operators.

In Section~\ref{Sec:GCU},
 we will show the bulk Higgs is preferable 
 for the accurate GCU,
 where  
 the favorite value of $1/(2R)$ 
 is about of ${\cal O}(10^{14})$ GeV. 
This seems dangerous for the proton decay 
through the dimension-six operators. 
Nevertheless, 
 this scenario does not have the dimension-six proton decay 
 processes, as the scenario
 in Section~\ref{Sec:brane-brane}.
Furthermore, this setup 
 can solve 
 the SUSY flavor problem 
 when the 
 hidden sector is 
 localized on the $y=0$ brane 
 via the gaugino mediation as in Section~\ref{Sec:brane-brane}. 
A difference here is that the bulk Higgs multiplets
 can also play a role of the SUSY breaking mediator 
 through the (flavor dependent) Yukawa interactions. 
As the gaugino mass, the SUSY-breaking masses of the Higgs fields,  
 $\tl{m}_h^2$, exist at the tree level via the contact interactions 
 $X^\dagger X H^\dagger H$.
These masses contribute to the flavor violation through the loop effects. 
Such contributions to the soft squared masses 
are evaluated in the leading-log approximation as
\begin{equation}
 \delta\tilde m^2 
 =2T Y^\dagger Y \tilde m_h^2 \frac{\ln(M_c/M_{\rm SUSY})}{16\pi^2},
\label{HiggsCont}
\end{equation}
 where $Y$ is the Yukawa matrix and $T$ is a group factor to be calculated 
 individually. 
The patterns of the flavor violations induced by 
 (\ref{HiggsCont}) are exactly
 the same as the well-known results in the MSSM 
 plus the right-handed neutrinos\cite{NuYUKAWA->LFV,Masiero:2002jn}
 with the universal SUSY breaking parameters
 at the cutoff scale, within the leading-log approximation.


In this case 
 the 
  $\mu$-term is generated through 
 $X^\dagger H_uH_d$
 with the same order as the soft SUSY masses, 
 $\mu \sim m_{3/2}\times(\dlt^2/\epsilon) \; (\sim M_{1/2})$ 
 because of the volume suppression factor
 in the interaction ($X^\dagger H_uH_d$) similar to 
 the gaugino masses. 
Therefore this scenario is 
 phenomenologically 
 favorable.\footnote{
We need a tuned coupling of $X^\dagger X H_uH_d$ to avoid
 a large $B$-parameter\cite{Kaplan:1999ac}.
The radion $F$-term                              
 might also solve the SUSY flavor problem,  
 however, the suitable $\mu$-term is not easily generated
 in the minimal field content as shown below. 
}
To be more precise, 
 since 
 the accurate GCU will require 
 $\delta \sim 1/32$ and 
 $\epsilon \sim 10^{-2}$,  
 which are read off from Eq.(\ref{4.17}), 
 the 
 soft SUSY masses and $\mu$ are smaller than 
 the gravitino mass as
 $0.1 \times m_{3/2}$. 

%


In the bulk Higgs scenario, 
 the $\mu$-term might be also obtained through 
 a non-canonical K\"{a}hler potential 
 ${\cal K}\ni H_uH_d+{\it h.c.}$ on 
 the branes and 
 a vanishing cosmological constant condition.  
This picks up   
 the SUSY and $R$-symmetry breaking
 effects\footnote{
The Peccei-Quinn (PQ) symmetry 
 is also broken by this non-canonical 
 K\"{a}hler potential.  
} in the supergravity (SUGRA) 
 setup,
 which is the so-called  
 Giudice-Masiero (GM) mechanism\cite{Giudice:1988yz}. 
It might 
 induce a small $\mu$-term as $\mu \sim m_{3/2}\times\dlt^2$,  
 while $M_{1/2}\sim m_{3/2}\times(\dlt^2/\epsilon)$ or 
 $M_{1/2}\sim m_{3/2}$ for the brane-localized hidden sector 
 or the radion $F$-term 
 scenario, respectively.\footnote{
In both cases the coupling between the gauge fields 
 and the hidden sector fields must be tuned to 
 be small to realize 
 $\mu \sim M_{1/2}$.
In such a case,  
 the anomaly mediation effects should be also 
 taken into account.
}


\subsection{Bulk Higgs and bulk matter}
\label{Sec:bulk-bulk}

The fourth scenario is putting both the 
 Higgs and matter fields in the bulk.  
This scenario guarantees 
 the charge quantization as well as
 the automatic anomaly cancellation 
 of $SO(10)$.\footnote{
The automatic anomaly cancellation is lost 
 if an $SO(10)$-incomplete multiplets are 
 put on the  $y=\pi R$ 
 brane. 
}  
As will be 
 shown in Section~\ref{Sec:GCU},
 the bulk Higgs setup is preferable 
 for the accurate GCU.

\subsubsection{Proton decay}

The dimension-five proton decay operators can be suppressed by the 
 approximate $R$-symmetry, 
 even though the triplet chiral partner $H_T^c$ 
 couples to the matter fields in this case. 
It should be noticed that 
 the triplet Higgs components~${H_T}$ 
 become super-heavy through their 
 $R$-symmetric 
 KK masses with the chiral partners, ${H_T^{c}}$, 
 instead of $R$-breaking 
 mixing masses between
 two $H_T$'s.  
This is an essence of the existence of 
 the (approximate) 
 $R$-symmetry, which prevents the
 dimension-five proton decay processes, 
 as keeping the triplet-doublet splitting\cite{Hall}. 
%
%
%
%
%

In order to suppress 
 the dimension-six proton decay processes, 
 the 1st and 2nd generations 
 should be localized on the $y=\pi R$
 brane, 
 as we have already shown in Section~\ref{Sec:brane-bulk}.
Let us examine how the proton stability constrains the localization of the 
 matter fields in more concrete. 
As discussed in Secion~\ref{Sec:brane-bulk}, 
 the localization of the $i$-th generation 
 is controlled by a kink mass $m_i$, 
  and we analyse the constraints on the parameters. 
The effective coupling (\ref{geff}) for the 1st generation 
 is constrained according to Eq.~(\ref{minMX}). 
For instance, a value $1/(2R)=3.6\times10^{14}{\rm GeV}$ which is 
 calculated in Secion~\ref{Sec:GCU} using the central values
 insists ${g_{\rm eff}^{\hat a}}/{g_4}<0.12$ for 
 $\alpha_H=0.01({\rm GeV})^3$ in order to be consistent with 
 $\tau_p(p\to e\pi)>1.6\times10^{33}{\rm years}$. 
This constraint is converted into that of the parameter $m_1$ through 
 Eq.~(\ref{geff}) and we find that $m_1R>13.6$.
This means that the 1st generation should be strictly localized 
 on the $y=\pi R$ brane, in practice. 

In addition, the localization of the 2nd generation is constrained 
 by another decay mode $\tau_p(p\to \nu_\mu K)>6.7\times10^{32}{\rm years}$, 
 which is induced through 
 $\overline{\bf16}_1{\bf16}_1\overline{\bf16}_2{\bf16}_2$. 
Now, the effective coupling is given as 
\begin{eqnarray}
 \sum_n\frac{g_{1,n}^{\hat a}}{g_4}\frac{g_{2,n}^{\hat a}}{g_4}
       \frac{1}{(2n+1)^2}&=&
 \frac\pi{32}
 \left\{
   \frac1{(m_1+m_2)R}-\frac{e^{-2\pi m_2R}}{m_1R}-\frac{e^{-2\pi m_1R}}{m_2R}
 \right. \nonumber\\
 &&\qquad\left.
  +e^{2\pi(m_1+m_2)R}
   \left(
    \frac1{m_1R}+\frac1{m_2R}-\frac1{(m_1+m_2)R}-2\pi
   \right)
 \right\} \nonumber\\
 &&\quad\times
 \brc{\coth (\pi m_1R)+1}\brc{\coth (\pi m_2R)+1},
\label{geff12}
\end{eqnarray}
 where $g_{i,n}^{\hat a}$ is defined by Eq.~(\ref{gn}) with replacing $m$ 
 by $m_i$ ($i=1,2$).
Assuming the same constraint~(\ref{minMX}) 
 also for this decay mode, 
 the square root of (\ref{geff12}) 
 is constrained to be smaller than $0.15$, leading to a constraint on $m_2$. 
For instance, we have $m_2R>4.0$ for $m_1R=13.6$.
For larger $m_1$, the constraint on $m_2$ becomes weaker.
In such a case, another constraint from the same decay mode induced by   
 $\bar{\bf16}_2{\bf16}_2\bar{\bf16}_2{\bf16}_2$ 
 may become dominant through the quark mixing. 
It constrains the effective coupling~(\ref{geff}) 
 with the replacement~$m\to m_2$. 
Then we obtain ${g_{\rm eff}^{\hat a}}/{g_4}<0.15\lmd^{-1}$, 
 where $\lmd$ is the mixing angle between the flavor and 
 the mass eigenstates. 
If it is given by the CKM mixing, {\it i.e.} $\lambda\sim0.22$, 
 we obtain $m_2R>0.5$.

In a similar way, the localization of the 3rd generation is possibly 
 constrained by a similar mode 
 $\tau_p(p\to \nu_\tau K)>6.7\times10^{32}{\rm years}$ 
 through the quark mixing between the 2nd and the 3rd generations. 
If the mixing is given by the CKM angle, {\it i.e.} $\lambda^2$, 
 the upper bound on the coupling is enhanced by 
 $\lambda^{-2}$ compared to that of the 2nd generation, leading to 
 no constraint on $m_3$.

\subsubsection{Yukawa interactions}

Due to the 5D ${\cal N}=1$ SUSY in the bulk, 
 the Yukawa interactions 
 cannot be written except on the branes. 

There are the following typical three cases 
 for the locations of the three generation matters. \\

\vspace{-4mm}
\noindent
{\bf Case A:} 
The 3rd generation is localized 
 around the $y=0$ brane.

In this case  
 there is a possibility to ensure the $SO(10)$ GUT 
 feature, \ie, the 
 $t$-$b$-$\tau$ 
 unification through
 the Yukawa interaction on the $y=0$ brane. 
The realistic Yukawa couplings for 
 the 1st and 2nd generations 
 are introduced  on the $y=\pi R$ brane 
 where the $SO(10)$ symmetry is broken down to the SM one. 
We must abandon the possibility to explain the fermion mass hierarchy 
 by the matter localization, and assume hierarchical couplings on the brane.

Because the off-diagonal terms in the Yukawa matrices on the $y=0$ brane 
 do not contribute to the CKM mixing due to the SU(2)$_R$ symmetry 
 in $SO(10)$, 
 the source of the mixing should be on the $y=\pi R$ brane. 
In order to reproduce the 2-3 mixing, the 3rd generation has to have 
 an overlapping with this brane no smaller than $\lambda^2$. 
This means that the $t$-$b$-$\tau$ unification is typically violated 
 by larger than $\lambda^4$.
\\

\vspace{-4mm}
\noindent       
{\bf Case B:} 
All generations are localized 
 around the $y=\pi R$ brane, 
 and the Yukawa interactions are also there.

This situation  is similar to the scenario 
 in Section 2.3, in which 
 the accurate GCU is realized as keeping the proton stability.
However, it looses both the explanation of the fermion mass hierarchy
 by their profiles and 
 the $t$-$b$-$\tau$ unification.  \\

\vspace{-4mm}
\noindent
{\bf Case C:} 
The 3rd generation is localized 
 around the $y=\pi R$ brane, and
 the 1st and 2nd generations are
 around the $y=0$ brane. 

In this case, the proton decays too rapidly 
 through the dimension-six processes, 
 which is enhanced for the accurate GCU. 
This difficulty can be avoided
  when the background geometry is warped. 
In the warped background\cite{Randall:1999ee}, 
 all the KK modes are localized around the $y=\pi R$ brane, and  
thus a mode localized around the $y=0$ brane has only a tiny overlap 
 with $A_\mu^{\hat a}$, which suppresses the proton decay.

We introduce Yukawa interactions with ${\cal O}(1)$ couplings 
 on the  $SO(10)$-breaking 
 $y=\pi R$ brane. 
In this case,   
although 
 the $t$-$b$-$\tau$ unification is 
 lost,  
 there is a possibility to explain the suitable fermion 
 mass hierarchies by the matter profiles\cite{ArkaniHamed:1999dc}.\\

%
%

\vspace{-4mm}
\subsubsection{SUSY breaking}
%
%
In general, 
 due to the existence of the matters in the bulk, 
 the SUSY flavor
 problem is not solved 
 unless another flavor-independent 
 SUSY-breaking mediation is introduced and becomes dominant. 
Now, the situation is better because the 1st and 2nd generations 
 are taken away from the $y=0$ brane to suppress the proton decay 
 via the dimension six operators. 
Thus, if the hidden sector where SUSY is broken is localized 
 on the $y=0$ brane, 
 the dangerous contact terms among the hidden sector and the 1st/2nd 
 generation are suppressed. 
For example, if we set $(m_1,m_2)R=(13.6,4.0)$ and 
 the 3rd generation localized around the $y=0$ brane, 
 the contact terms for the scalar soft masses 
 of the 1st and 2nd generations are exponentially suppressed as 
\begin{equation}
 \delta\tilde m^2\sim
 \left(
 \begin{array}{ccc}
 10^{-37}&10^{-24}&10^{-18} \\
 10^{-24}&10^{-11}&10^{- 5} \\ 
 10^{-18}&10^{- 5}&1 
 \end{array}
 \right)\tilde m_0^2.
\end{equation} 
Thus, we can conclude that non-negligible contact terms can appear only 
 in the $(3,3)$ element, 
 in the flavor basis.
In order to evaluate the flavor violation, 
 we have to move to the mass basis. 
In the case when the mixing is given by the CKM matrix, 
 the tree level off-diagonal elements are given as 
\begin{equation}
 \delta\tilde m^2\sim
 \left(
 \begin{array}{ccc}
           & \lambda^5 & \lambda^3 \\
 \lambda^5 &           & \lambda^2 \\ 
 \lambda^3 & \lambda^2 & 
 \end{array}
 \right)\tilde m_0^2.
\label{FV}
\end{equation} 

The diagonal elements are generated through the gaugino loop 
 as Eq.~(\ref{GauginoCont}) in the leading-log approximation.
Thus, assuming $\tilde m_0\sim M_{1/2}$, we can see 
 that the off-diagonal elements (\ref{FV}) give 
 interesting predictions just around 
 the present bounds, calculated in Ref.\cite{ConstraintFV} 
 for $M_{\rm SUSY}\sim 350{\rm GeV}$ and not so large $\tan\beta$. 
Now, $\tan\beta$ is large to realize the $t$-$b$-$\tau$ unification, 
 and thus the bounds cannot be applied as they are in the reference. 
Nevertheless, this observation is useful to get a rough sketch 
 whether the contact terms are crucially dangerous or not.
The actual bounds in this model would be revealed by a more detailed 
 analysis using the full RGEs, 
 which is one of our future works.

As for 
 the $\mu$-term, the situation is the 
 same as Section 2.3, where
 the direct interaction 
 $X^\dagger H_u H_d$ on the brane works well. 
Also, the scalar masses of the Higgs fields 
 exist via $X^\dagger X H^\dagger H$, 
 and contribute to the flavor violation through the loop effects 
 as evaluated in Eq.~(\ref{HiggsCont}), 
 giving a similar contributions as in the MSSM plus the right-handed 
 neutrinos with the universal soft terms.

\section{Gauge Coupling Unification}
\label{Sec:GCU}

Since higher dimensional gauge theories are non-renormalizable, 
 it is not easy to trace the flow of each gauge coupling constant
 above the compactification scale. 
Nevertheless it is known that, if there is the unified symmetry 
 in the bulk, flows of differences of two different gauge coupling 
 constants, $\delta\alpha_i^{-1}\equiv \alpha_i^{-1}-\alpha_1^{-1}$, 
 are at most logarithmic in the orbifold models\cite{Hall,Hall:2001xb,Nomura:2001mf}. 
Thus we can examine whether the three
 gauge couplings are unified or not.
Essentially the same discussion can be also applied to the iGUTs,  
 and we show it in the following.
%

It is convenient to introduce the following non-analytic but continuous 
 function for $x\geq1$:
\begin{equation}
 f(x) = \sum_{k=1}^{k=k_x-1}(-1)^{k}\ln\left(\frac{k+1}k\right) 
        +(-1)^{k_x}\ln\left(\frac{x}k_x\right), 
\end{equation}
 where $k_x$ is the natural 
  number that satisfies $x-1\leq k_x<x$.
This function converges for large $x$ as 
 $f(\infty)=-\ln(\pi/2)\sim-0.45$. 
In the following analysis, 
 we approximate this function by $f(\infty)$ for 
 $x\gtrsim10$, because   
 an error induced by this approximation is
 of ${\cal O}(1/(2x))$. 

First,  
 let us evaluate the contributions from the bulk ${\bf 10_H}$ 
 Higgs hypermultiplet 
 with BCs 
 in Eq.(\ref{HiggsBC}) 
 above the (half of) compactification scale.   
Here we do not introduce 
 parity-odd bulk masses, for simplicity. 
The KK spectra of the doublets and triplets are 
 $n/R$ and $(n+1/2)/R$, respectively. 
A pair of the doublet and triplet 
 compose a (full) multiplet of $SU(5)$,
 and then the contribution from this pair is common to the flow 
 of each coupling.
This means that 
  a triplet contributes 
 to the flows of $\dlt\alp_i^{-1}$ 
 by the same factor as a doublet but with 
 the opposite sign. 
In each KK state, 
 there are four doublets or four triplets 
 except for the zero-modes.
(The zero-modes consist of only the two Higgs doublets.)
Therefore the contribution of the ${\bf 10_H}$ hypermultiplet 
 above $1/(2R)$ is given by 
\begin{equation}
 \Delta_H \delta\alpha_i^{-1}(\mu) 
 = -2\frac{\delta C_D^i}{2\pi} f\left({2R}\mu\right)
 \,\sim\, \frac{\delta C_D^i}{\pi}\ln\left(\frac\pi2\right), 
\label{BulkHiggs}
\end{equation}
 where $\delta C_D=(0,1/5,-3/10)$ is the contribution by the Higgs doublet 
 to the flow of $\delta\alpha_i^{-1}$. 
On the other hand, the contribution from
 the two Higgs-doublet superfields 
 localized on the brane is given by 
\begin{equation}
 \Delta_H \delta\alpha_i^{-1}(\mu) 
 = -\frac{\delta C_D^i}{\pi} \ln\left({2R}\mu\right), 
\label{BraneHiggs}
\end{equation}
 as in the usual 4D models.
Comparing Eqs.(\ref{BulkHiggs}) and (\ref{BraneHiggs}), 
 we notice that 
 the sign is flipped 
 when the Higgs multiplets start propagating in the bulk. 

Next we evaluate the contributions from
 the gauge multiplet with the BCs 
 in Eq.(\ref{gaugeBC}). 
The KK modes with a mass~$n/R$ in $A_\mu^{a}$ 
 compose an $SO(10)$ multiplet 
 together with those  
 with a mass $(n+1/2)/R$ in $A_\mu^{\hat a}$, so that
 the former contributes to the flows of $\dlt\alp_i^{-1}$ 
 by the same 
 factor as the latter but with the opposite sign. 
Since a massless vector (chiral) supermultiplet contributes 
 $-3$ ($1$), 
 a massive vector supermultiplet contributes  $-3+1=-2$.  
Thus,
 above $1/(2R)$,
 the $SO(10)$ multiplet~$(A_\mu^a,A_\mu^{\hat a})$ 
 contributes to the flow of $\dlt\alp_i^{-1}$ as
\begin{equation}
 \Delta_g \delta\alpha_i^{-1}(\mu)  = 
 -\frac{\delta C_g^i}{2\pi}
   \left(-2\ln\left({2R}\mu\right)-f\left({2R}\mu\right)\right)
 \,\sim\, -\frac{\delta C_g^i}{2\pi} 
          \left(-2\ln\left({2R}\mu\right)+\ln\left(\frac\pi2\right)\right), 
\end{equation}
 where $\delta C_g=(0,2,3)$ is the contribution from the 
 MSSM gauge sector. 

As for the matter fields, they do not contribute to the flows of 
 $\dlt\alp_i^{-1}$ because they compose
 degenerate 
 $SO(10)$ full multiplets.
Then,
 in summary, 
 we obtain
\begin{equation}
 \delta\alpha_i^{-1}(\mu) \sim 
 \delta\alpha_i^{-1}\left(\frac1{2R}\right)
 -\frac{\delta C_g^i}{2\pi} 
  \left(-2\ln\left({2R}\mu\right)+\ln\left(\frac\pi2\right)\right)
 +\Delta_H \delta\alpha_i^{-1}(\mu).
\end{equation}
Defining $\Lg$ by $\delta\alpha_2^{-1}(\Lg)=0$
 in the MSSM, 
 the value of $\alpha_i^{-1}\left(1/(2R)\right)$ 
 is determined as
\begin{equation}
  \delta\alpha_i^{-1}\left(\frac1{2R}\right)
 =\delta\alpha_i^{-1}(\Lg)
 +\frac{\delta b_i}{2\pi}\ln\left(2R\Lg\right) ,
\end{equation}
where 
 $\delta b_i=(0,-28/5,-48/5)$ is 
 the difference of the beta functions  
 in the MSSM.

Now  
 we can estimate the deviations from the MSSM, 
 depending on the Higgs profiles.  
We determine the value of 
 $\Lc$ by use of $\delta\alpha_2^{-1}(\Lc)=0$. 
By 
 imposing $\delta\alpha_3^{-1}(\Lc)=0$,
 we determine 
 $1/({2R})$ and $\Lc$ as a function 
 of $\delta\alpha_3^{-1}(\Lg)$, 
 which is the problematic disagreement of
 the QCD coupling 
 in the 4D minimal $SU(5)$ GUT. 
Neglecting 
 the GUT threshold correction, 
 the deviation 
 is estimated as
 $\delta\alpha_3^{-1}(\Lg)=0.855\pm0.315$\cite{Hall:2001xb}.  
Anyhow, 
 the GCU 
 crucially depends on  
 whether the Higgs fields are located in 
 the bulk or on the brane, so  
 we analyze the GCU in each case.

\subsection*{\bf Bulk Higgs case:}

\noindent
Equation (\ref{BulkHiggs}) derives 
\begin{equation}
 \delta\alpha_i^{-1}(\mu) \sim 
 \delta\alpha_i^{-1}\left(\frac1{2R}\right)
 -\frac{\delta C_g^i}{2\pi} 
  \left(-2\ln\left({2R}\mu\right)+\ln\left(\frac\pi2\right)\right)
 +\frac{\delta C_D^i}{\pi}\ln\left(\frac\pi2\right).
\end{equation}
Taking $\delta\alpha_i^{-1}(\Lc)=0$,
 we can calculate 
 $(\ln\left({2R}\Lg\right), 
 \ln\left({2R\Lc}\right))$ as 
\begin{equation}
 \left(\begin{array}{l}
 \ln\left({2R}\Lg\right) \\
 \ln\left({2R\Lc}\right)
 \end{array}\right)
 =
 \left(\begin{array}{l}
 \frac13{2\pi}\delta\alpha^{-1}_3(\Lg)-\ln\left(\frac{\pi}2\right)\\
 \frac7{15}{2\pi}\delta\alpha^{-1}_3(\Lg)-\ln\left(\frac{\pi}2\right)
 \end{array}\right)
 =
 \left(\begin{array}{l}
 4.00 \\ 5.78
 \end{array}\right)
\end{equation}
 for $\delta\alpha_3^{-1}(\Lg)=0.855$.
This means 
\begin{equation}
 (2R\Lg,\,\,2R\Lc)=(55,\,\,320),
\label{4.17}
\end{equation}
which is consistent with
 Refs.\cite{Hall:2001xb}. 
In this case, the mass of the lightest modes in $A_\mu^{\hat a}$ is 
 evaluated as $1/(2R)=3.6\times10^{14}$~GeV, which is 
 too light to be consistent with
 the proton decay constraint 
 unless  the coupling
 $g_{\rm eff}^{\hat a}$ is small as  
 $g_{\rm eff}^{\hat a}/g_4 <$ 
 $0.12$ $(0.21)$ for $\alpha_H=0.01$ $(0.003)$.\footnote{
For the smaller value by 1-$\sigma$, 
 $\delta\alpha_3^{-1}(\Lg)=0.539$, the gauge boson mass is
 modified as 
 $1/(2R)=1.9\times10^{15}$GeV, which still requires 
 a little bit small $g_{\rm eff}^{\hat a}$ or $\abs{\alpha_H}$ as 
 $g_{\rm eff}^{\hat a}/g_4 <$ 
 $0.60$ $(1.1)$ for $\alpha_H=0.01$ $(0.003)$. 
}
It can be achieved when 
 the 1st generation matter is localized around $y=\pi R$.  
(A typical case is $g_{\rm eff}^{\hat a}=0$ which  corresponds to 
 the matters strictly localized on 
 the $y=\pi R$ brane.)
This constraint 
 plays a crucial role for the construction
 of models as shown in Section 2.4.

\subsection*{\bf Brane Higgs case:}


\noindent
By similar calculations, 
 we obtain 
\begin{equation}
 \left(\begin{array}{l}
 \ln\left({2R}\Lg\right) \\
 \ln\left({2R\Lc}\right)
 \end{array}\right)
 =
 \left(\begin{array}{l}
 -8.5\\-13
 \end{array}\right)
\end{equation}
by use of Eq.(\ref{BraneHiggs}). 
However, 
 this means 
 $\Lc<\Lg<1/(2R)$, 
 a nonsense relation. 
This implies 
 the precise GCU is difficult
 in the brane Higgs scenario. 
Thus, 
 introduction of 
 extra $SO(10)$ incomplete 
 multiplets on the $y=\pi R$ brane 
 might be required for the precise GCU.

\vspace{2mm}

Recalling that 
 the light triplet Higgs multiplets are 
 preferred for the GCU in the 4D minimal GUT\cite{Murayama:2001ur},
 the bulk Higgs scenario, which 
 can have the light triplets, 
 might be preferred than the brane Higgs scenario. 
(We should emphasize again that 
 the light triplets in the 4D GUT induces too rapid proton decay.)

\section{Interval BCs by $\bdm{f}\!\bdm{ake}$ Higgs}
\label{Sec:BC}

Some of the interval BCs can be obtained from an orbifold~$S^1/Z_2$ 
 by a method which we call the {\it fake} Higgs construction. 
In this paper we focus on such types of BCs,
 which are expected to be consistent 
 with the tree-level unitarity and the Ward-Takahashi
 identities\cite{Csaki:2003dt,Uekusa}. 
The {\it fake} Higgs construction of the interval BCs was first 
 introduced in Ref.\cite{Nomura:2001mf}. 
For reader's convenience, we review this method in this section. 
We discuss general arguments first, and then give 
 the $SO(10)$ BCs on an interval.

\subsection{General arguments}

In the orbifold, BCs are strictly restricted 
 by  the orbifolding parity 
 if there are no boundary terms. 
Namely, fields with even (odd) parities follow 
 the Neumann (Dirichlet) BCs automatically. 
However, in the interval, 
 the even (odd) parity does not
 automatically correspond to 
 the Neumann (Dirichlet) BC. 
Thus, more general BCs are possible 
 on the interval, which 
 broaden the possibility 
 of the model-building. 
Some of them are obtained by introducing 4D scalar fields 
 on the boundaries,
 whose VEVs break part of the residual symmetries 
 under the orbifold projection, and taking their VEVs to infinity. 
We name such boundary fields as {\it fake} Higgs fields
 because they are not dynamical degrees of freedom 
 after taking the limit. 
The effects of the boundary Higgs fields
 are replaced 
 by the boundary 
 masses after they get VEVs. 
The detailed calculations are provided in Appendices. 
In this subsection we will explicitly see how the boundary masses 
 change the mass spectra and BCs of the bulk fields 
 in some simple examples 
 to illustrate the situation.

\subsubsection{Gauge sector}

Here we consider a case that part of the gauge symmetries is 
 broken at $y=\pi R$ 
 by the boundary masses~$\cM_{\hat{a}}$ 
 for the gauge fields~$A^{\hat{a}}_\mu$, 
 which are induced by the VEVs of the boundary {\it fake}
 Higgs fields. 
The mass spectrum is determined by Eq.(\ref{spctrm:gauge}) 
 in Appendix~\ref{gauge_sector}. 
In the flat spacetime, it becomes 
\be
 \tan(m_{\hat{a},n}\pi R) = \frac{\cM_{\hat{a}}}{2m_{\hat{a},n}}, 
 \label{eq:tan}
\ee
and the mode functions (profiles of wave functions) are given by
\be
 f^{\hat{a}}_n(y) = \brkt{\frac{\pi R}{2}
 +\frac{\cM_{\hat{a}}}{4m_{\hat{a},n}^2+\cM_{\hat{a}}^2}}^{-1/2}
 \cos(m_{\hat{a},n}y), 
 \label{md_eq:gauge_flat}
\ee
where $m_{\hat{a},n}$ are solutions of Eq.(\ref{eq:tan}).

In the case of no boundary mass, \ie, $\cM_{\hat{a}}=0$, 
 the gauge field~$A^{\hat{a}}_\mu$ follows the Neumann BCs 
 at both boundaries and the mass eigenvalues are 
 $m_{\hat{a},n}=n/R$ ($n$: integer),
 which is just the case of the orbifold. 
If we turn on the boundary mass~$\cM_{\hat{a}}$, 
 the eigenvalues are shifted as 
\be
 m_{\hat{a},n}=
 \frac{n}{R}+\frac{1}{\pi R}
 \arctan\brkt{\frac{\cM_{\hat{a}}}{2m_{\hat{a},n}}}. 
\label{200}
\ee 
For a finite $\cM_{\hat{a}}$, the shift of the mass eigenvalue 
 monotonically decreases as the KK level~$n$ increases, and 
 becomes negligible for $m_{\hat{a},n}\gg\cM_{\hat{a}}$. 
In the limit of $\cM_{\hat{a}}\to\infty$, on the other hand, 
 all eigenvalues are uniformly shifted by $1/(2R)$, 
 which indicates that the boundary condition at $y=\pi R$ 
 changes from Neumann to Dirichlet.
This can be seen explicitly from Eq.(\ref{md_eq:gauge_flat}). 
The boundary value of the
 mode function at $y=\pi R$ is given by 
\be
 \abs{f^{\hat{a}}_n(\pi R)} = 
 \brkt{\frac{\pi R}{2}+\frac{\cM_{\hat{a}}}{4m_{\hat{a},n}^2
 +\cM_{\hat{a}}^2}}^{-1/2}
 \abs{\frac{2m_{\hat{a},n}}{\sqrt{4m_{\hat{a},n}^2+\cM_{\hat{a}}^2}}}, 
\ee
by using Eq.(\ref{eq:tan}). 
In the limit of $\cM_{\hat{a}}\to\infty$, this goes down to zero, 
 \ie, $f^{\hat{a}}_n(y)$ follows the Dirichlet BC at $y=\pi R$. 
We should remember that
 the parity eigenvalues never change 
 at any BCs realized by the {\it fake} Higgs.

\subsubsection{Hypermultiplet sector}

Next we see mass spectra of hypermultiplets 
 in the presence of boundary masses. 
In Appendix~\ref{Bmt}, we consider a case that 
the bulk hypermultiplets have mass terms localized at $y=\pi R$. 
Here let us focus on a case that two hypermultiplets have only 
the boundary Dirac mass~$\eta$ 
 in a flat spacetime,\footnote{The 
 parameter~$\eta$ is dimensionless, which corresponds to 
 a ratio of the {\it fake} 
 Higgs VEV to the 5D cutoff scale~$\Lc$. 
} 
for simplicity. 
We take the orbifold parities of the hypermultiplets as 
 Eq.(\ref{Z2_parity1}). 
In this case, 
 Eq.(\ref{detM1}) is reduced to 
\be
 \tan^2(m_n\pi R) = \abs{\eta}^2,  \label{eq:tan2}
\ee
where $m_n$ is a mass eigenvalue of the $n$-th KK mode. 
The solution of Eq.(\ref{eq:tan2}) is given by  
\be
 m_n = \frac{n}{R}\pm\frac{\arctan\abs{\eta}}{\pi R}. 
 \label{m_n1:fm}
\ee
It should be noticed that all mass eigenvalues receive
 the same shift due to the
 boundary mass~$\eta$ independently of the KK level~$n$, 
 even for finite $\eta$. 
This is in contrast to the case of the gauge sector
 in Eq.(\ref{200}). 
In the limit of $\abs{\eta}\to\infty$, 
 the shift of the mass eigenvalues becomes $1/(2R)$, 
 which means that BC of even-parity fields at $y=\pi R$ 
 changes from Neumann to Dirichlet. 
The mode functions defined in Eq.(\ref{KK_expand:fm}) are 
 (for $0<y<\pi R$) given as 
\bea
 f_{h,n}(y) \eql \alp_{h,n}\cos(m_n y), \;\;\;
 f_{H,n}(y) = \pm\frac{\eta^*}{\abs{\eta}}\alp_{h,n}^*\cos(m_n y), \nonumber\\
 f_{h,n}^c(y) \eql -\alp_{h,n}^*\sin(m_n y), \;\;\;
 f_{H,n}^c(y) = \mp\frac{\eta}{\abs{\eta}}\alp_{h,n}\sin(m_n y), 
 \label{explicit_mdfct1}
\eea
where the double signs correspond to that in Eq.(\ref{m_n1:fm}), 
 and the complex constants~$\alp_{h,n}$'s are determined by 
 the normalization condition, Eq.(\ref{norm_cond2}). 
Again, we should remember that
 the parity eigenvalues do not change 
 even when BCs change.

When we take 
 orbifold parities 
 as Eq.(\ref{Z2_parity3}), 
 Eq.(\ref{eq:tan2}) is modified as 
\be
 \cot^2(m_n\pi R) = \abs{\eta}^2, 
\ee
where 
 the mass spectrum is given by 
\be
 m_n = \frac{n+\frac{1}{2}}{R}\pm\frac{\arctan\abs{\eta}}{\pi R}. 
 \label{m_n2:fm}
\ee
It means that 
 the shift of the eigenvalues by $\eta$ is the same as 
 that in Eq.(\ref{m_n1:fm}). 
Notice that 
 no zero-mode exists when $\eta=0$, 
 however, 
 it appears 
 in the limit of $\abs{\eta}\to\infty$. 
This indicates that BC of $h$ at $y=\pi R$ changes from
 Dirichlet to Neumann. 
The mode functions are the same as Eq.(\ref{explicit_mdfct1}),  
 but $m_n$ in the arguments are now given by Eq.(\ref{m_n2:fm}) 
 and the double signs correspond to that in it.

Note that BCs of the mode functions~$f_{\phi,n}$ and $f_{\phi,n}^c$ 
 ($\phi=h,H$) are related to each other 
 by the bulk mode equations in Eq.(\ref{bulk:md_eq1}). 
Therefore if BC of a chiral multiplet changes from Neumann to Dirichlet, 
 that of its chiral partner inevitably changes 
 from Dirichlet to Neumann.
Since the orbifold 
 parities are unchanged by the boundary terms,  
 the mode function of a parity-odd field becomes 
 discontinuous at the boundary when BC
 changes from Dirichlet to Neumann.   

A similar relation exists 
 between $V^A$ and $\Phi^A$ ($A=a,\hat{a}$) 
 in the gauge multiplet. 
As will be mentioned in Appendix~\ref{gauge_sector}, 
 the gauge-scalar multiplet~$\Phi^A$ is absorbed into 
 the $\cN=1$ vector multiplet~$\der_y V^a$. 
This means that the mode functions for the former are 
 the same as the derivative of the mode functions for the latter. 
On the other hand,
 the mode equation 
 for $A_y^A$ is decoupled from $A_\mu^A$ 
 by choosing a particular gauge-fixing function. 
So the mode functions and KK spectrum for $A_y^A$ are independent 
 of the boundary masses for $A_\mu^A$. 
 ({\it See, for example}, Ref.\cite{HS}.)
It seems contradict with the above relation between 
 $V^A$ and $\Phi^A$. 
However, we should remind 
 that $A_y^A$ is unphysical degree of freedom 
 because it is eaten by $A_\mu^A$
 through the ``Higgs mechanism''. 
In fact the mode function and the spectrum for $A_y^A$ are 
 gauge-dependent. 
Thus, we can always choose a gauge-fixing function for 
 the KK spectrum of $A_y^A$
 to coincide with that of $A_\mu^A$ 
 (except for the zero-mode). 
The $\cN=1$ superfield description in this paper 
 corresponds to this gauge. 

When we take orbifold parities 
 as Eq.(\ref{Z2_parity2}), 
 situation is quite different from
 the previous two cases. 
Equation (\ref{detM2}) is reduced to 
\be
 \sin(m_n\pi R)\cos(m_n\pi R) = 0, 
\ee
which means that the spectrum~$m_n=n/(2R)$ is unchanged 
by the boundary mass~$\eta$. 
Thus the zero-mode always exists irrespectively
 of the value of $\eta$. 
The mode functions have explicit $\abs{\eta}$-dependence 
 (for $0<y<\pi R$) as  
\bea
 f_{h,n}(y) \eql \alp_{h,n}\cos(m_n y), \;\;\;
 f_{H,n}(y) = \alp_{H,n}\cos(m_n y), \nonumber\\
 f_{h,n}^c(y) \eql -\alp_{h,n}^*\sin(m_n y), \;\;\;
 f_{H,n}^c(y) = -\alp_{H,n}^*\sin(m_n y), 
\eea
where 
\be
 \alp_{H,n} = \begin{cases} -\frac{1}{\eta}\alp_{h,n}^* & (\mbox{$n$: even}), \\ 
 \eta^*\alp_{h,n}^* & (\mbox{$n$: odd}). \end{cases} 
 \label{rel:alphas}
\ee
This is in contrast to the previous cases, where 
 $\abs{\eta}$-dependence of the mode function 
 appears only through the mass eigenvalue. 
For $\sin(m_n\pi R)=0$ (\ie, $n$ is even), for example, 
 the modes reside only in $(H,H^c)$ when $\eta=0$. 
Equation (\ref{rel:alphas}) means that 
 this mode continuously moves from $(H,H^c)$ to $(h,h^c)$ 
 as $\abs{\eta}$ increases. 
We can also infer this
 behavior from the fact that BCs are interchanged 
 between the two hypermultiplets when $\abs{\eta}$ goes from zero 
 to infinity.

Finally we consider a mixing mass 
 between a bulk hypermultiplet~$(H,H^c)$  
 and a chiral multiplet~$\chi$ localized on the $y=\pi R$ brane. 
Here we focus on a simple case that a bulk mass term is absent 
and the spacetime is flat. 
Then Eq.(\ref{det_cM2}) is reduced to
\be
 \tan(m_n\pi R) = \frac{\abs{\xi}^2}{2(m_n\pm\abs{m_\chi})}, 
 \label{eq:tan3}
\ee
for the parity assignment of Eq.(\ref{Z2_parity4}), and 
\be
 \cot(m_n\pi R) = -\frac{\abs{\xi}^2}{2(m_n\pm\abs{m_\chi})}, 
\ee
for the parity assignment of Eq.(\ref{Z2_parity5}). 
The mixing parameter~$\xi$ has mass-dimension~$1/2$ and 
the mass parameter for $\chi$, $m_\chi$, has mass-dimension~$1$. 
({\it See} Eq.(\ref{cL:mix1}).)
Equation (\ref{eq:tan3}) has the same forms as Eq.(\ref{eq:tan}) 
 if we replace $\abs{\xi}^2$ with $\cM_{\hat a}$ and 
 set $m_\chi=0$. 
Thus the $\abs{\xi}$-dependence of the spectrum is similar 
to that of the gauge multiplet. 
Due to the existence of the boundary term at $y=\pi R$, 
the parity-odd field becomes discontinuous there. 
{}From Eq.(\ref{rel:Hc-phi}) 
 (or the counterpart in the case of Eq.(\ref{Z2_parity5})), 
 the 4D chiral multiplet~$\chi$ is 
 expressed as this discontinuity.

\subsection{$\bdm{F}\!\bdm{ake}$ Higgs in $\bdm{SO(10)}$ GUT}
\label{Sec:FakeHiggs}

In Section~\ref{Sec:BCbyHand} we introduced $SO(10)$ incomplete
 multiplets and $SO(10)$-breaking interactions on the $y=\pi R$ brane 
 by hand, since the gauge group is already reduced to 
 the SM gauge symmetry there. 
In this subsection,
 we show an explicit realization
 of this setup 
 by the {\it fake} Higgs construction. 
We start from an $SO(10)$-invariant
 theory on $S^1/Z_2$, where 
 $A_y$ has odd-parity  
 so that it has no zero-modes.  
This means that the charge quantization and anomaly cancellation 
 are ensured 
in this setup. 


In order to obtain the BCs in Eq.(\ref{gaugeBC}), 
 we put ${\bf 45_H}$, ${\bf 16_H}$, and
 ${\bf \overline{16}_H}$ 
 {\it fake} Higgs fields 
 on the $y=\pi R$ brane.
The ${\bf 45_H}$ Higgs takes a VEV of    
 diag.$(\sigma_2, \sigma_2, \sigma_2, 0, 0)\ v_{\bf 45}$,
 which reduces the gauge symmetry as
 $SO(10) \rightarrow 
 SU(3)_c \times SU(2)_L \times SU(2)_R \times U(1)_{B-L}$
 at the energy scale of $v_{\bf 45}$, where $\sigma_2$ is a Pauli matrix. 
And the ${\bf 16_H}$ and ${\bf \overline{16}_H}$ 
 take VEVs in a $D$-flat direction 
 which lead to the breaking of 
 $SU(2)_R \times U(1)_{B-L} \rightarrow U(1)_Y$. 
Taking the VEVs of 
 the {\it fake} Higgs 
 to infinity, the BCs for 
 the gauge multiplet in Eq.(\ref{gaugeBC}) 
 are obtained. 
The ${\bf 45_H}$, ${\bf 16_H}$ and ${\bf \overline{16}_H}$  
 {\it fake} Higgs 
 fields are assumed to have suitable 
 interactions among them       
 in order not to leave 
 light (colored) degrees of freedom.\footnote{
 If these interactions 
 are absent, components with $(3,2)_{1/6}$ and $(3^*,2)_{-2/3}$ 
 for $SU(3)_c\times SU(2)_L\times U(1)_Y$ 
 become pseudo-NG bosons 
 even in the limit of infinite VEVs.
 (If gauge interactions are switched off, they become
 exact NG bosons.)} 



\subsubsection{Triplet-Doublet Splitting}

Realization of the triplet-doublet splitting 
 can be achieved by using a technique of 
 Dimopoulos-Wilczek (DW) mechanism\cite{DW}. 
The VEV of the ${\bf 45_H}$ {\it fake} Higgs 
 in a direction of $U(1)_{B-L}$ generator
 induces the triplet Higgs masses as keeping 
 the doublet Higgs massless.
It is justified 
 as far as the doublet Higgs fields are contained
 in ${\bf 10_H}$, 
 since the doublets in ${\bf 10_H}$ have vanishing 
 $U(1)_{B-L}$ charges. 
Here
 we have to introduce additional ${\bf 10'_H}$ on the $y=\pi R$ brane 
 to allow the coupling between ${\bf 10_H}$ and ${\bf 45_H}$.
This is because two identical
 ${\bf 10_H}$ multiplets cannot form
 Yukawa interactions with ${\bf 45_H}$ according to  
 the $SO(10)$ group structure, 
\begin{equation}
 {\bf 10}\times{\bf 10}={\bf 1}_S+{\bf 45}_A+{\bf 54}_S, \nonumber 
\end{equation}
 where subscript $S$ ($A$) indicates that 
 the product is 
 (anti-)symmetric. 
The brane superpotential which realizes 
 the triplet-doublet splitting 
 is given by\footnote{Here we assume that 
 terms such as ${\bf 10_H}^2$ which destroy the 
 DW mechanism are absent.} 
\begin{equation}
 W_{DW}=\delta(y-\pi R)\left(
           y_{DW}\frac{{\bf 10_H}}{\sqrt{\Lc}^{\delta_{\bf 10_H}}}
                 {\bf 45_H}{\bf 10'_H} +
           {m_{DW}}{\bf 10'_H}^2
           \right) ,
\end{equation}
 where 
 $\delta_{\bf 10_H}=1(0)$ for bulk (brane) ${\bf 10_H}$ field. 
It is natural to regard ${\bf 10'_H}$ 
 as a {\it fake} Higgs too, so that
 $m_{DW}$ should be taken to infinity.          
When $\bf 10_H$ is a brane field, 
 only the MSSM doublet Higgs components remain to be massless 
 and other fields decouple by getting super-heavy 
 with large masses $y_{DW}v_{\bf 45}$ and $m_{DW}$. 
When $\bf 10_H$ is a bulk field, the KK spectrum is 
 given as Eq.(\ref{eq:tan3}) by identifying 
 $\xi$ and $m_\chi$ 
 with $y_{DW}\vev{{\bf 45_H}}/\sqrt{\Lc}$ and $m_{DW}$,
  respectively. 
Thus, for the triplet components, the lightest KK modes obtain 
 masses of $\cO(1/(2R))$, while  
 the doublet components remain to be massless, 
 which do not couple to ${\bf 45_H}$. 


\subsubsection{Brane Interactions}

Here we comment on an idea of taking
 zero limits of
 the {\it fake} Higgs couplings in order
 to obtain the finite matter 
 interactions and masses 
 effectively. 
We know
 that
  the wrong GUT relations of the mass spectra between the down-type quarks
 and charged leptons can be modified by the effects of
 $SU(5)$-breaking
 VEVs. 
The realistic Yukawa matrices might be
 induced from the brane interactions,
\begin{equation}
 \delta(y-\pi R)Y_{n,m}
   \frac{\bf 16}{\sqrt{\Lc}^{\delta_{\bf 16}}}
   \frac{\bf 16}{\sqrt{\Lc}^{\delta_{\bf 16}}}
   \frac{\bf 10_H}{\sqrt{\Lc}^{\delta_{\bf 10_H}}}
   \left(\frac{\bf 45_H}{\Lc}\right)^n
   \left(\frac{\bf 16_H\overline{16}_H}{\Lc^2}\right)^m,
\label{21}
\end{equation}
 where $\dlt_{\bf 16}=1(0)$ for
 bulk (brane) ${\bf 16}$ matter.
The lowest order, $n=m=0$, gives an
 $SO(10)$-symmetric Yukawa coupling.\footnote{
If ${\bf16_H}$ and ${\bf \overline{16}_H}$ do not couple
 to the matter fields,
 which means
 $m=0$, the CKM mixing angles vanish because of the $SU(2)_R$
 symmetry which commutes with ${\bf \langle 45_H\rangle}$.}
This is an example of the FN mechanism.
The effective Yukawa couplings are divergent by the
 infinite VEVs of the {\it fake} Higgs for
 the finite magnitude of
 $Y_{n,m}$ ($n,m>0$).
So, in order to obtain finite Yukawa couplings from
 Eq.(\ref{21}),
 the couplings $Y_{n,m}$ must be infinitely small
 as keeping $Y_{n,m}{\bf \langle 45_H\rangle}^n
 ({\bf \langle 16_H\rangle \langle\overline{16}_H\rangle})^m$ finite.
The finite Majorana masses of
 the right-handed neutrinos might be
 also obtained from the brane interaction,
\begin{equation}
 \delta(y-\pi R)\omg
 \frac{\bf 16}{\sqrt{\Lc}^{\delta_{\bf 16}}}
 \frac{\bf 16}{\sqrt{\Lc}^{\delta_{\bf 16}}}
 \frac{{\bf \overline{16}_H}{\bf \overline{16}_H}}{\Lc},
\end{equation}
where a coupling $\omg$
 should be tuned
 for 
 the suitable magnitudes of Majorana masses.\footnote{
The KK masses
 do not break the lepton number.}   
%
%
\ignore{
Anyhow, in the {\it fake} Higgs construction,
 we cannot avoid this kind of fine-tuning
 in order to obtain the finite matter
 interactions and masses
 of $SO(10)$-breaking.\footnote{
Notice that, however, if we regard the {\it fake} Higgs construction 
 as a technical method just for ensure the unitarity, charge quantization 
 anomaly cancellation and so on, we need not mind the fine-tuning 
 in the {\it fake} Higgs sector.
}}

\section{Summary and discussion} \label{summary}

We have discussed 5D SUSY GUTs on the interval, where 
 the gauge multiplets propagate in the 5D bulk. 
Interval BCs make the rank reduction
 of the gauge symmetry possible  
 in contrast to the orbifold BCs. 
Although this idea of the rank reduction by BCs 
 is well-known\cite{Csaki:2003dt}, most models use it to break 
 the electro-weak symmetry\cite{Csaki:2003zu} 
 but the application to the GUT breaking has not 
 been studied 
 except for the trinification model\cite{Carone:2004rp}. 

We have investigated the 5D $SO(10)$ iGUT, in which 
 the gauge symmetry is directly reduced to 
 the SM without introducing GUT-breaking Higgs fields. 
This is in contrast to the orbifold GUTs where 
the rank reduction is impossible. 
We can also consider iGUTs based on other higher-rank gauge symmetries, 
 such as $E_6$. 

To be more concrete, we investigated the GCU, the proton decay and
 the $SO(10)$ features such as $t$-$b$-$\tau$ unification and 
 charge quantization for different localization 
 of the matter and Higgs fields. 
We also estimated the flavor violations by the SUSY partners.
We briefly summarize our results:  
\vspace{-1mm}
\begin{enumerate}
\item Bulk Higgs scenario: \\
The GCU is improved, \ie,   
 the small disagreement of the QCD coupling from the 
 predicted value in the 4D GCU 
 can be corrected by the existence of 
 the light triplet Higgs modes. 
For this purpose, a compactification scale 
 lower than the GUT scale is required,  
 demanding the matter fields 
 to be localized around the $y=\pi R$ brane
 for the proton stability.
\vspace{-2mm}
 
Because the bulk Higgs fields can couple to the SUSY breaking 
 sector, the $mu$ term can be induced through a contact term. 
In a similar way, the scalar soft squared masses for the Higgs fields 
 can be generated and then induces the flavor violations via the 
 RGE effects, which is similar to that in the MSSM with the 
 right-handed neutrinos.

\item Brane Higgs scenario: \\
We can introduce only the doublet components of the physical
 Higgs on the  $SO(10)$-breaking brane. 
This means that there is no 
 dimension-five 
 proton decay operators induced by 
 the triplet Higgses. 
Additional $SO(10)$-incomplete multiplets might be 
 needed for realizing the precise GCU. 
\vspace{-2mm}

In order to realize an appropriate $\mu$ term, 
 some additional mechanism such as the NMSSM may be required.

\item Bulk matter scenario:  \\
The charge quantization of $Q(p^+)=-Q(e^-)$ is ensured. 
If the 3rd generation matter field is localized around 
 the $SO(10)$-preserving brane, 
 the $t$-$b$-$\tau$ unification can be also realized. 
\vspace{-2mm}

Since bulk matters in general cause the SUSY flavor 
 problem, another source of SUSY breaking 
 may be needed which induces 
 flavor-independent soft masses. 
When the 1st and 2nd generations are localized around the 
 $y=\pi R$ brane, which is required by the proton decay 
 constraint for the improved GCU, the flavor violations 
 are suppressed. 
When the 3rd generation has overlapping with the 
 SUSY breaking brane, the contact term generates 
 a sizable contribution to the $(3,3)$ element 
 of the scalar soft mass matrices at the mediation scale, 
 in the flavor basis. 
Although the off-diagonal elements are negligible in the flavor basis, 
 the flavor violation can occur 
 through the mixing matrix between the flavor and the mass bases. 
Especially if this mixing matrix is given by the CKM matrix, 
 the flavor violation is estimated around the experimental bounds. 


\item Brane matter scenario:  \\
We loose some of the GUT-predictions such as  
 the charge quantization and  
 the $t$-$b$-$\tau$ unification. 
\vspace{-2mm}

The SUSY flavor problem can be solved 
 by the sequestering (gaugino mediation), and  
 the dimension-six proton decay processes 
 are absent in this setup. 

\end{enumerate}
\vspace{-2mm}
In each case, all dimension-five proton decay processes
 can be suppressed by
 the (approximate) $R$-symmetry\cite{Hall}. 
The realistic Yukawa interactions and the Majorana masses 
 can be reproduced with the help of the superpotential 
 localized on the $SO(10)$-breaking brane. 
As for the anomaly cancellation,
 the automatic cancellation of $SO(10)$ is lost  once  $SO(10)$-incomplete
 multiplets are introduced 
 on the  $SO(10)$-breaking brane. 
Here, we would emphasize that the couplings between the bulk matter 
 fields and the gauge fields for the broken generators 
 ({\it e.g.} the $X$ gauge boson for SU(5) models) are non-vanishing, 
 and induce the proton decay via the dimension-six operators.
This is in great contrast with the orbifold GUTs where 
 these couplings are absent because of the constrained 
 parity assignments. 

The interval BCs were first considered in Ref.\cite{Csaki:2003dt}. 
Then, Ref.\cite{Uekusa} investigates their consistency and 
 finds some BCs that violate the tree-level unitarity 
 and the Ward-Takahashi identities. 
In order to avoid such dangerous BCs, we used BCs obtained 
 by introducing Higgs fields localized on the boundaries 
 and taking a limit that their VEVs go to infinity\cite{Csaki:2003dt}, 
 which we call the {\it fake} Higgs construction.

Finally, let us comment on the warped spacetime. 
The iGUTs can be applied
 also in the warped 5D background\cite{Randall:1999ee}. 
The equations in the Appendices are useful also in the warped setup. 
We mentioned that 
 the constraints from the dimension-six proton decay 
 are largely modified from the flat case 
 due to the wave-function profiles 
 of the lower KK modes, 
 while the discussion on the symmetry breaking pattern 
 and the location of the hidden sector are not. 
As for the GCU, 
 we have a technical difficulty in the analysis 
 since the KK mass spectrum cannot
 be calculated analytically, 
 although it is expected that 
 qualitative features are
 not drastically changed from the flat case. 
If 
 the gauge coupling evolution is defined by  
 two-point Green functions of the gauge fields 
 with external lines on the UV brane, 
 it develops logarithmically, 
 and thus is calculable\cite{Randall:2001gc}. 
In this case, 
 the difference of the gauge couplings are
 frozen out above the IR scale.
Namely the GCU is the 
 same as the situation in the MSSM, 
 when the IR scale is the GUT scale.

\vspace{1cm}
\leftline{\bf Acknowledgments}

We would like to thank Y. Hosotani
 for useful discussions.  
N.~H. is supported by the Grant-in-Aid for Scientific Research, Ministry
 of Education, Science and Culture, Japan (No.16540258 and No.17740146). 
Y.~S. is supported by the Japan Society for the 
 Promotion of Science for Young Scientists (No.0509241). 
T.~Y. is supported in part by The 21st Century COE Program 
 ``Towards a New Basic Science; Depth and Synthesis''.

\vspace{1cm}

\appendix

\section{KK expansion with boundary masses} \label{KKexpand_bdm}

In this appendix, 
 we review derivations of the KK spectra and 
 profiles in a general setup
 with boundary masses. 
The 5D metric is given by 
\be
 ds^2 = G_{MN}dx^M dx^N = e^{-2\sgm(y)}\eta_{\mu\nu}+dy^2, 
\ee
where $M,N=0,1,2,3,5$ are 5D indices, $\mu,\nu=0,1,2,3$ are 4D ones, 
 and $y\equiv x^5$. 
The warp factor~$\sgm(y)$ is 
 assumed to be a monotonic and nondecreasing function of $y$ 
 and $\sgm(0)=0$.

\subsection{Gauge sector} \label{gauge_sector}

A 5D gauge multiplet consists of 
 a gauge-scalar~$\Sgm$, a gauge field~$A_M$,
 and gauginos~$\lmd^i$,  
 where $i=1,2$ is the $SU(2)_R$ index. 
Each field is matrix-valued, \ie, 
\be
 A_M = \sum_A A_M^AT^A, 
\ee
where $T^A$ is a generator of the gauge group. 
The 5D Lagrangian is written in the
 4D  $\cN=1$ superspace by introducing 
the following $\cN=1$ superfields\cite{AGW,MartiPomarol},\footnote{
For the extension to the 5D SUGRA, see Ref.\cite{AS}. 
We will follow the notation of Ref.\cite{WB}.} 
\bea
 V \defa -\tht\sgm^\mu\bar{\tht}A_\mu
 +ie^{\frac{3}{2}\sgm}\tht^2\bar{\tht}\bar{\lmd}^1
 -ie^{\frac{3}{2}\sgm}\bar{\tht}^2\tht\lmd^1
 +\frac{1}{2}\tht^2\bar{\tht}^2D, \nonumber\\
 \Phi \defa \frac{1}{2}\brkt{\Sgm+iA_y}+e^{\frac{1}{2}\sgm}\tht\lmd^2
 +\tht^2 F_\Phi, 
\eea
where $D$ and $F_\Phi$ are auxiliary fields. 
We focus on a simple case that the orbifold projection does not 
break the gauge group at all. 
Namely, the orbifold parity is assigned as 
\be
 V\;(+,+), \;\;\;
 \Phi\;(-,-). \label{Z2_parity:gauge}
\ee
The left (right) signs denote the parities at $y=0$ ($y=\pi R$).

The 5D Lagrangian is expressed as 
\bea
 \cL^{\rm gauge} \eql 
 \sbk{\int d^2\tht\;\frac{1}{2g_5^2}\tr\brkt{\cW^\alp \cW_\alp}+\hc} 
 +e^{-2\sgm}\int d^4\tht\;\frac{1}{g_5^2}\tr\brkt{\cV_5^2}, 
\eea
where $g_5$ is the 5D gauge coupling, 
$\cW_\alp$ and $\cV_5$ are the gauge-covariant quantities defined as 
\bea
 \cW_\alp \defa \frac{1}{4}\bar{D}^2 e^{V}D_\alp e^{-V} 
 = -\frac{1}{4}\bar{D}^2D_\alp V+\cdots, \nonumber\\
 \cV_5 \defa e^V\der_ye^{-V}+\Phi+e^V\Phi^\dagger e^{-V}
 = -\der_yV+\Phi+\Phi^\dagger+\cdots, 
\eea
where the ellipses denote quadratic and higher terms. 
The normalization of the generators are taken as 
\be
 \tr(T^AT^B)=\frac{1}{2}\dlt^{AB}. 
\ee

We introduce 4D chiral multiplets~$\phi_0^I$ and $\phi_\pi^J$ 
 localized at the orbifold boundaries at~$y=0$ and $\pi R$, respectively. 
The indices $I,J$ run over different irreducible representations 
of the gauge group. 
They interact with the 5D gauge multiplet as 
\bea
 \cL^{\rm bd} \eql e^{-2\sgm}\int d^4\tht\;\brc{\sum_I \phi_0^{I\dagger}
 e^{-V}\phi_0^I \dlt(y)
 +\sum_J \phi_\pi^{J\dagger}e^{-V}\phi_\pi^J \dlt(y-\pi R)}+\cdots, 
\eea
where the ellipsis denotes the self-interaction terms of 
 $\phi_{0,\pi}$. 

The above Lagrangians are invariant under
 the (super-) gauge transformation, 
\bea
 e^V &\to& e^\Lmd e^Ve^{\Lmd^\dagger}, \nonumber\\
 \Phi &\to& e^\Lmd\brkt{\Phi-\der_y}e^{-\Lmd}, \nonumber\\
 \phi_0^I &\to& e^{\Lmd(0)}\phi_0^I, \;\;\;\;\;
 \phi_\pi^J \to e^{\Lmd(\pi R)}\phi_\pi^J.  
 \label{gauge_trf}
\eea
The transformation parameter~$\Lmd$ is a chiral superfield. 
Under this transformation, the gauge-covariant
 quantities transform as 
\bea
 \cW_\alp &\to& e^\Lmd\cW_\alp e^{-\Lmd}, \nonumber\\
 \cV_5 &\to& e^\Lmd\cV_5 e^{-\Lmd}. 
\eea
By choosing the gauge parameter~$\Lmd$ as 
\be
 \exp\brc{\Lmd(x,y)} = \cP\exp\brc{-\int_0^y dy'\;\Phi(x,y')}, 
\ee
we move into the gauge where $\Phi=0$. 
The symbol~$\cP$ stands for the path ordering operator from
 left to right. 
Recall that all (non-zero)
 KK modes of $A_y$ are absorbed into those of 
 $A_\mu$ by the ``Higgs mechanism'',
 and the latter obtain the KK masses. 
Thus we can call it unitary gauge. 
Note that $V$ is no longer in the Wess-Zumino gauge, 
 and its lowest and the next lowest components for $\tht,\bar{\tht}$ 
 are physical degrees of freedom. 

Now we assume that the scalar components of $\phi_0^I$ and $\phi_\pi^J$ 
 get VEVs and break the gauge group to a
 subgroup at the boundaries. 
Then the Lagrangian becomes 
\bea
 \cL \eql 
 \sbk{\int d^2\tht\;\frac{1}{2g_5^2}\tr\brkt{\cW^\alp \cW_\alp}+\hc} 
 +\frac{e^{-2\sgm}}{g_5^2}\int d^4\tht\;\tr\brc{(-\der_y V)^2+\cdots}
 \nonumber\\
 &&+\frac{e^{-2\sgm}}{2g_5^2}\int d^4\tht\;
 \brc{\sum_{I,A,B}\cM_0^{(I)AB}V^AV^B\dlt(y)
 +\sum_{J,A,B}\cM_\pi^{(J)AB}V^AV^B\dlt(y-\pi R)+\cdots}, \nonumber\\
 \label{L_gauge}
\eea
where $\cM_0^{(I)AB}$ and $\cM_\pi^{(J)AB}$ are the boundary mass 
parameters defined by 
\be
 \cM_0^{(I)AB}\equiv 
 g_5^2\vev{\phi_0^I}^\dagger T^AT^B\vev{\phi_0^I}, \;\;\;\;\;
 \cM_\pi^{(J)AB}\equiv 
 g_5^2\vev{\phi_\pi^J}^\dagger T^AT^B\vev{\phi_\pi^J}. 
\ee
The ellipses in Eq.(\ref{L_gauge}) shows the 
 terms involving the fluctuation 
 around the VEVs, which decouple in the limit of
 $\cM_{0,\pi}\to\infty$. 
Here we have assumed that SUSY is preserved when $\phi_{0,\pi}$ 
 get the VEVs. 

In the following discussions,
 we consider a case of $\cM_0^{(I)AB}=0$. 
Then we can always diagonalize the matrix~$\sum_J\cM_\pi^{(J)AB}$ 
 for the indices~$A,B$ by using the gauge symmetry, 
 \ie, $\sum_J\cM_\pi^{(J)AB}=\cM_A\dlt^{AB}$. 
Thus Eq.(\ref{L_gauge}) becomes 
\bea
 \cL \eql 
 \sbk{\frac{1}{4g_5^2}\int d^2\tht\;\sum_A\cW^{A\alp} \cW_\alp^A+\hc} 
 \nonumber\\
 &&-\frac{1}{2g_5^2}\int d^4\tht\;
 \sum_A V^A\brc{\der_y\brkt{e^{-2\sgm}\der_y V^A}
 -e^{-2\sgm}\cM_AV^A\dlt(y-\pi R)+\cdots}, 
 \label{L_gauge2}
\eea
where we have performed the partial integration. 
Now we expand the 5D superfield~$V$ into 4D KK modes,  
\be
 V^A(x,y,\tht,\bar{\tht}) = \sum_n f_n^A(y)V_n^A(x,\tht,\bar{\tht}). 
\ee
The mode equation for the KK modes is read off from Eq.(\ref{L_gauge2}) as    
\be
 \der_y\brkt{e^{-2\sgm}\der_y f_n^A}
 -e^{-2\sgm}\cM_A f_n^A\dlt(y-\pi R) = -m_{A,n}^2f_n^A.  
 \label{md_eq:gauge}
\ee
Since $V^A$ is a $Z_2$-even superfield, 
the mode functions~$f^A_n(y)$ are even functions around 
$y=0,\pi R$. 
Thus from Eq.(\ref{md_eq:gauge}), we obtain the following BCs. 
\bea
 \der_y f^A_n|_{y=0} \eql 0, \nonumber\\
 \sbk{\der_y f^A_n}_{\pi R-\ep}^{\pi R+\ep} \eql \cM_Af_n^A(\pi R). 
 \label{bdcd:gauge}
\eea
The first condition is the ordinary Neumann BC   
 while the second one is a mixed-type BC. 
In fact the latter is reduced to the Neumann BC 
 in the limit of $\cM_A\to 0$, 
 and it becomes the Dirichlet BC 
 in the limit of $\cM_A\to\infty$.

The general solution of Eq.(\ref{md_eq:gauge}) is written as 
\be
 f^A_n(y) = \alp^A_n C(y,m_{a,n})+\bt^A_n S(y,m_{a,n}), 
 \label{gen_sol:gauge}
\ee
where $\alp^A_n$ and $\bt^A_n$ are real constants
 determined 
 by the BCs. 
The functions~$C(y,m)$ and $S(y,m)$ are defined 
 in Appendix~\ref{bases_fcn}. 
The first condition in Eq.(\ref{bdcd:gauge}) means $\bt^A_n=0$. 
So the second condition is translated into 
\be
 -2C'(\pi R,m_{a,n}) = \cM_A C(\pi R,m_{a,n}), 
 \label{spctrm:gauge}
\ee
 where the prime denotes the $y$-derivative. 
This determines the mass spectrum~$\brc{m_{a,n}}$. 
The remaining constant~$\alp^A_n$ is fixed by 
 the normalization condition, 
\be
 \int_0^{\pi R}dy\;\brkt{f^A_n(y)}^2 = 1. 
\ee

\subsection{Hypermultiplet sector} \label{matter_sector}

Next we consider a matter sector 
 of hypermultiplets~$(\bdm{H}_i,\bdm{H}^c_i)$, 
 where $\bdm{H}_i$ and $\bdm{H}^c_i$ are 
 $\cN=1$ chiral superfields and  
 belong to conjugate representations   
 of the gauge group. 
Namely, under the gauge transformation Eq.(\ref{gauge_trf}), 
 they transform as 
\be
 \bdm{H}_i \to e^\Lmd\bdm{H}_i, \;\;\;\;\;
 \bdm{H}^c_i \to \bdm{H}_i^c e^{-\Lmd}. 
\ee
The index~$i$ runs over the irreducible representations. 
The bulk Lagrangian of this sector is given by 
\bea
 \cL^{\rm hyper}_{\rm bulk} \eql e^{-2\sgm}\int d^4\tht\;
 \sum_i\brkt{\bdm{H}^\dagger_ie^{-V}\bdm{H}_i
 +\bdm{H}^c_i e^{V}\bdm{H}^{c\dagger}_i} 
 \nonumber\\
 &&+e^{-3\sgm}\left[\int d^2\tht\;\sum_i\left\{
 \frac{1}{2}\brkt{\bdm{H}^c_i\der_y\bdm{H}_i
 -\der_y\bdm{H}^c_i\bdm{H}_i} \right.\right. \nonumber\\
 &&\hspace{35mm}\left.\left. \rule[-2mm]{0mm}{8mm}
 -\bdm{H}^c_i\Phi\bdm{H}_i
 +M_i\vep(y)\bdm{H}^c_i\bdm{H}_i\right\}+\hc\right], 
\eea
where $M_i$'s are bulk mass parameters and 
$\vep(y)$ is the periodic step function. 

For simplicity, 
 let us focus on two hypermultiplets  among
 $(\bdm{H}_i,\bdm{H}^c_i)$,
 and denote them as 
 $(h,h^c)$ and $(H,H^c)$. 
They are components either in the same gauge multiplet or 
 different one. 
Then the Lagrangian for them is written as 
\bea
 \cL^{\rm hyper}_{\rm bulk} \eql e^{-2\sgm}\int d^4\tht\brc{
 \abs{h}^2+\abs{h^c}^2+\abs{H}^2+\abs{H^c}^2} \nonumber\\
 &&+e^{-3\sgm}\left[\int d^2\tht\left\{
 \frac{1}{2}\brkt{h^c\der_y h-h\der_y h^c+H^c\der_y H-H\der_y H^c} 
 \right.\right. \nonumber\\
 &&\hspace{30mm}\left.\left.\rule[-2mm]{0mm}{8mm}
 +M_h\vep(y) h^ch+M_H\vep(y) H^cH \right\}+\hc \right]+\cdots, 
 \label{L_bulk}
\eea
where $M_h$ and $M_H$ are the bulk mass parameters. 
In the case that $(h,h^c)$ and $(H,H^c)$ belong to the same 
 gauge multiplet,
 $M_h=M_H$. 

\subsubsection{Boundary mass terms} \label{Bmt}

Here we consider effects from mass terms 
 localized at $y=\pi R$, 
 which are induced by the VEVs of $\phi_\pi^J$.
(We do not consider the boundary masses coming from
 $\phi_0^I$, for simplicity.) 
Each chiral superfield has an opposite orbifold parity 
 to the chiral partner (contained in the same hypermultiplet). 
Thus there are the following three cases 
 according to the orbifold parity assignments.

\begin{flushleft}
\bf Case 1
\end{flushleft}
First we consider a case that both the hypermultiplets 
have the same parities at both the orbifold boundaries, \ie, 
\be
 h\;(+,+), \;\;\; h^c\;(-,-), \;\;\;
 H\;(+,+), \;\;\; H^c\;(-,-). \label{Z2_parity1}
\ee
The left (right) signs denote the parities at $y=0$ ($y=\pi R$).
In this case the most general boundary mass terms are given by 
\bea
 \cL^{\rm hyper}_{\rm bd} \eql e^{-3\sgm}\left[\int d^2\tht\;
 \bpx h & H \epx \bpx \kp & \eta \\
 \eta & \lmd \epx \bpx h \\ H \epx \dlt(y-\pi R)+\hc\right], 
\eea
where the Majorana masses~$\kp$, $\lmd$ and 
 the Dirac mass~$\eta$ are dimensionless
 parameters.\footnote{These parameters correspond
 to ratios of the boundary Higgs VEVs 
 to the 5D cutoff scale~$\Lc$.}
We treat them as complex parameters, although  
 two phases among $\kp$, $\lmd$ and $\eta$ 
 can be absorbed by field redefinitions unless 
 the phases of $h$ and $H$ are fixed in another sector.

The mode equations are given by 
\bea
 &&-\frac{1}{4}\bar{D}^2\bar{h}-e^{-\sgm}\brkt{\der_y-\frac{3}{2}\sgm'
 -M_h\vep}h^c+2e^{-\sgm}(\kp h+\eta H)\dlt(y-\pi R) = 0, 
 \nonumber\\
 &&-\frac{1}{4}\bar{D}^2\bar{h}^c+e^{-\sgm}\brkt{\der_y-\frac{3}{2}\sgm'
 +M_h\vep}h =0, \nonumber\\
 &&-\frac{1}{4}\bar{D}^2\bar{H}-e^{-\sgm}\brkt{\der_y-\frac{3}{2}\sgm'
 -M_H\vep}H^c+2e^{-\sgm}(\eta h+\lmd H)\dlt(y-\pi R) = 0, \nonumber\\
 &&-\frac{1}{4}\bar{D}^2\bar{H}^c+e^{-\sgm}\brkt{\der_y-\frac{3}{2}\sgm'
 +M_H\vep}H = 0. 
\label{57}
\eea
The BCs at $y=0$ are determined 
 only by the orbifold parities\footnote{
The BCs for $h$ and $H$ do not 
 provide independent informations from those for $h^c$ and $H^c$,     
 because the former mode functions are 
 related to the latter ones through
 the bulk equations of motion. ({\it See} Eq.(\ref{bulk:md_eq1}).)
} as   
\be
 \left. h^c \right|_{y=0} = \left. H^c \right|_{y=0} = 0, 
\ee
since there are no boundary terms there. 
On the other hand, the 
 BCs at $y=\pi R$ are modified 
 by the boundary mass terms as 
\bea
 \sbk{h^c}^{\pi R+\ep}_{\pi R-\ep} \eql 
 2\brc{\kp h+\eta H}_{y=\pi R}, \nonumber\\
 \sbk{H^c}^{\pi R+\ep}_{\pi R-\ep} \eql 
 2\brc{\eta h+\lmd H}_{y=\pi R}. 
 \label{sf_bdcd1}
\eea
Now we expand the 5D superfields into 4D KK modes as 
\bea
 h(x,y,\tht) \eql \sum_n e^{\frac{3}{2}\sgm}f_{h,n}(y)h_n(x,\tht), 
 \;\;\;\;\;
 h^c(x,y,\tht) = \sum_n e^{\frac{3}{2}\sgm}f_{h,n}^c(y)h^c_n(x,\tht), 
 \nonumber\\
 H(x,y,\tht) \eql \sum_n e^{\frac{3}{2}\sgm}f_{H,n}(y)H_n(x,\tht), 
 \;\;\;\;\;
 H^c(x,y,\tht) = \sum_n e^{\frac{3}{2}\sgm}f_{H,n}^c(y)H^c_n(x,\tht). 
\label{KK_expand:fm}
\eea
The mode equations in the bulk ($0<y<\pi R$) are given by
\bea
 \brkt{\der_y+M_\phi}f_{\phi,n} \eql m_n e^\sgm f_{\phi,n}^{c*}, 
 \nonumber\\
 \brkt{\der_y-M_\phi}f_{\phi,n}^c \eql -m_n e^\sgm f_{\phi,n}^*, 
 \label{bulk:md_eq1}
\eea
where $\phi=h,H$. 
%
{}From the BCs in Eq.(\ref{sf_bdcd1}), the mode functions 
 must satisfy the 
 conditions,  
\bea
 f_{h,n}^c(0) \eql f_{H,n}^c(0) = 0, 
 \label{BC_0} \\
 f_{h,n}^c(\pi R-\ep) \eql 
 -\kp f_{h,n}(\pi R)-\eta f_{H,n}(\pi R), \nonumber\\
 f_{H,n}^c(\pi R-\ep) \eql 
 -\eta f_{h,n}(\pi R)-\lmd f_{H,n}(\pi R). 
 \label{mf_bdcd1}
\eea
Solutions of Eq.(\ref{bulk:md_eq1}) with the BC in Eq.(\ref{BC_0}) 
 are given by 
\bea
 f_{\phi,n}(y) \eql \alp_{\phi,n}e^{-M_\phi y}C_{M_\phi}(y,m_n), 
 \nonumber\\
 f_{\phi,n}^c(y) \eql -\alp_{\phi,n}^*e^{M_\phi y}S_{-M_\phi}(y,m_n), 
 \label{gen_sol1}
\eea
where $\phi=h,H$, and 
 $(\alp_{h,n},\alp_{H,n})$ are complex constants 
 determined by the BCs. 
The functions~$C_M(y,m)$ and $S_M(y,m)$ are defined 
 in Appendix~\ref{bases_fcn}. 
Thus the BCs in Eq.(\ref{mf_bdcd1})
 are rewritten as 
\be
 \cM_4 \bpx \Re \alp_{h,n} \\ \Re \alp_{H,n} \\ 
 \Im \alp_{h,n} \\ \Im \alp_{H,n} \epx = 0, 
 \label{AB_cond1}
\ee
with 
\be
 \cM_4 \equiv \bpx \kp_R\tl{C}_{M_h}-\tl{S}_{-M_h} & \eta_R\tl{C}_{M_H} 
 & -\kp_I\tl{C}_{M_h} & -\eta_I\tl{C}_{M_H} \\ 
 \eta_R\tl{C}_{M_h} & \lmd_R\tl{C}_{M_H}-\tl{S}_{-M_H} 
 & -\eta_I\tl{C}_{M_h} & -\lmd_I\tl{C}_{M_H} \\ 
 \kp_I\tl{C}_{M_h} & \eta_I\tl{C}_{M_H} 
 & \kp_R\tl{C}_{M_h}+\tl{S}_{-M_h} & \eta_R\tl{C}_{M_H} \\
 \eta_I\tl{C}_{M_h} & \lmd_I\tl{C}_{M_H} & 
 \eta_R\tl{C}_{M_h} & \lmd_R\tl{C}_{M_H}+\tl{S}_{-M_H}
 \epx, \label{expr_M1}
\ee
where 
\be
 \tl{S}_{-M} \equiv e^{M\pi R}S_{-M}(\pi R,m_n), \;\;\;\;\;
 \tl{C}_M \equiv e^{-M\pi R}C_M(\pi R,m_n), 
\ee
and $\kp_R\equiv\Re\kp$, $\kp_I\equiv\Im\kp$, 
 and so on. 
The condition that 
 Eq.(\ref{AB_cond1}) has 
 a nontrivial solution is 
\be
 \det\cM_4 = \tl{C}_{M_h}^2\tl{C}_{M_H}^2\brkt{ 
 \tl{T}_{M_h}^2\tl{T}_{M_H}^2-2\abs{\eta}^2\tl{T}_{M_h}\tl{T}_{M_H}
 -\abs{\lmd}^2\tl{T}_{M_h}^2
 -\abs{\kp}^2\tl{T}_{M_H}^2+\abs{\kp\lmd-\eta^2}^2} = 0, 
 \label{detM1}
\ee
where
\be
 \tl{T}_M \equiv \frac{\tl{S}_{-M}}{\tl{C}_M}. \label{def_tlT}
\ee
Equation (\ref{detM1}) determines the mass spectrum. 
The complex constants~$\alp_{h,n}$ and $\alp_{H,n}$ are determined by 
 Eq.(\ref{AB_cond1}) with the solution of Eq.(\ref{detM1}),
 and 
 the normalization condition, 
\be
 \int_0^{\pi R}dy\;\brc{\abs{f_{h,n}(y)}^2+\abs{f_{H,n}(y)}^2}
 = 1. \label{norm_cond2}
\ee

\begin{flushleft}
\bf Case 2
\end{flushleft}
Next we consider a case that one hypermultiplet has the same 
 parities at both boundaries while the other has opposite 
 parities, \ie, 
\be
 h\;(+,-), \;\;\; h^c\;(-,+), \;\;\;
 H\;(+,+), \;\;\; H^c\;(-,-). \label{Z2_parity2}
\ee
In this case the most general boundary mass terms are given by 
\bea
 \cL^{\rm hyper}_{\rm bd} \eql e^{-3\sgm}\left[\int d^2\tht\;
 \bpx h^c & H \epx \bpx \kp & \eta \\
 \eta & \lmd \epx \bpx h^c \\ H \epx \dlt(y-\pi R)+\hc\right], 
\eea
where the dimensionless mass parameters~$\kp$, $\lmd$ and $\eta$ 
 are complex. 
Through similar calculations to the Case~1, we obtain 
 an equation that determines the mass spectrum. 
It corresponds to 
 Eq.(\ref{detM1}) with the replacement of 
\be
 (\tl{S}_{-M_h},\tl{C}_{M_h}) \to 
 (\tl{C}_{M_h},-\tl{S}_{-M_h}). 
\ee
Namely, the mass spectrum is determined by  
\be
 \det\cM_4 = \tl{S}_{-M_h}^2\tl{C}_{M_H}^2\brkt{
 \tl{T}_{M_h}^{-2}\tl{T}_{M_H}^2+2\abs{\eta}^2\tl{T}_{M_h}^{-1}\tl{T}_{M_H}
 -\abs{\lmd}^2\tl{T}_{M_h}^{-2}
 -\abs{\kp}^2\tl{T}_{M_H}^2+\abs{\kp\lmd-\eta^2}^2} = 0.  
 \label{detM2}
\ee

\begin{flushleft}
\bf Case 3
\end{flushleft}
Finally we consider a case that both the hypermultiplets have opposite 
parities at the two boundaries, \ie, 
\be
 h\;(+,-), \;\;\; h^c\;(-,+), \;\;\;
 H\;(+,-), \;\;\; H^c\;(-,+). \label{Z2_parity3}
\ee
In this case the most general boundary mass terms are given by 
\bea
 \cL^{\rm hyper}_{\rm bd} \eql e^{-3\sgm}\left[\int d^2\tht\;
 \bpx h^c & H^c \epx \bpx \kp & \eta \\
 \eta & \lmd \epx \bpx h^c \\ H^c \epx \dlt(y-\pi R)+\hc\right], 
\eea
where the dimensionless mass parameters~$\kp$, $\lmd$ and $\eta$ 
are complex. 
The equation that determines the mass spectrum 
 corresponds to  
 Eq.(\ref{detM1}) with the replacement,  
\bea
 (\tl{S}_{-M_h},\tl{C}_{M_h}) &\to& 
 (\tl{C}_{M_h},-\tl{S}_{-M_h}), \nonumber\\
 (\tl{S}_{-M_H},\tl{C}_{M_H}) &\to& 
 (\tl{C}_{M_H},-\tl{S}_{-M_H}). 
\eea
Thus, the mass spectrum is determined by  
\be
 \det\cM_4 = \tl{S}_{-M_h}^2\tl{S}_{-M_H}^2\brkt{
 \tl{T}_{M_h}^{-2}\tl{T}_{M_H}^{-2}
 -2\abs{\eta}^2\tl{T}_{M_h}^{-1}\tl{T}_{M_H}^{-1}
 -\abs{\lmd}^2\tl{T}_{M_h}^{-2}
 -\abs{\kp}^2\tl{T}_{M_H}^{-2}+\abs{\kp\lmd-\eta^2}^2} = 0. 
 \label{detM3}
\ee

\subsubsection{Mixing with boundary fields}

Now 
 let us consider effects from brane mass terms 
 between a bulk hypermultiplet~$(H,H^c)$ and 
 a 4D chiral superfield~$\chi$ localized on the $y=\pi R$ brane. 

\begin{flushleft}
\bf Case 4
\end{flushleft}
First we consider a case that the hypermultiplet has 
the same parities at both boundaries, \ie, 
\be
 H\;(+,+), \;\;\; H^c\;(-,-). \label{Z2_parity4}
\ee 
The boundary Lagrangian in this case is 
\be
 \cL_{\rm bd} = \brc{e^{-2\sgm(\pi R)}\int d^4\tht\; \abs{\chi}^2
 +e^{-3\sgm(\pi R)}\sbk{\int d^2\tht\;
 \brkt{\xi H\chi+\frac{1}{2}m_\chi\chi^2}+\hc}}\dlt(y-\pi R). 
 \label{cL:mix1}
\ee
The constants~$\xi$ and $m_\chi$ have 
mass-dimension~$1/2$ and $1$, respectively. 
The equations of motion are 
\bea
 &&-\frac{1}{4}\bar{D}^2\bar{H}+e^{-\sgm}
 \brc{-\brkt{\der_y-\frac{3}{2}\sgm'-M_H\vep}H^c 
 +\xi\chi\dlt(y-\pi R)} = 0, \nonumber\\
 &&-\frac{1}{4}\bar{D}^2\bar{H}^c
 +e^{-\sgm}\brkt{\der_y-\frac{3}{2}\sgm'
 +M_H\vep}H = 0, \nonumber\\
 &&-\frac{1}{4}\bar{D}^2\bar{\chi}
 +e^{-\sgm(\pi R)}\brc{\xi H|_{y=\pi R}+m_\chi\chi}
 = 0, 
 \label{EOM2}
\eea
From the first equation, we obtain a relation 
 between the boundary and bulk superfields as 
\be
 \sbk{H^c}^{\pi R+\ep}_{\pi R-\ep} = \xi\chi. 
 \label{rel:Hc-phi}
\ee
Using this relation, the last equation in Eq.(\ref{EOM2}) is 
rewritten as 
\be
 \sbk{-\frac{1}{4}\bar{D}^2\bar{H}^c
 -e^{-\sgm}\brkt{\frac{\abs{\xi}^2}{2}H
 -\hat{m}_\chi H^c}}_{y=\pi R-\ep} = 0, 
\ee
where $\hat{m}_\chi\equiv m_\chi\bar{\xi}/\xi$. 
Thus the mode functions satisfy the following BC, 
\be
 \frac{\abs{\xi}^2}{2}f_{H,n}(\pi R) -\hat{m}_\chi f^c_{H,n}(\pi R-\ep) 
 = -e^{\sgm(\pi R)} m_n f^{c*}_{H,n}(\pi R-\ep). 
\ee
This is translated by using Eq.(\ref{gen_sol1}) into  
\be
 \brkt{\frac{\abs{\xi}^2}{2}\tl{C}_{M_H}-e^{\sgm(\pi R)}m_n\tl{S}_{-M_H}}
 \alp_{H,n}+\hat{m}_\chi\tl{S}_{-M_H}\alp_{H,n}^* = 0. 
\ee
The condition for it to have a nontrivial solution is 
\be
 \brkt{\frac{\abs{\xi}^2}{2}\tl{C}_{M_H}-e^{\sgm(\pi R)}m_n\tl{S}_{-M_H}}^2
 -\abs{\hat{m}_\chi\tl{S}_{-M_H}}^2 = 0,  
\ee
or
\be
 \tl{T}_{M_H} = 
 \frac{\abs{\xi}^2}{2\brkt{e^{\sgm(\pi R)}m_n\pm\abs{m_\chi}}}. 
 \label{det_cM2}
\ee

\begin{flushleft}
\bf Case 5
\end{flushleft}
Next we consider a case that the hypermultiplet has 
opposite parities at both boundaries, \ie, 
\be
 H\;(+,-), \;\;\; H^c\;(-,+). \label{Z2_parity5}
\ee
The boundary Lagrangians in this case are 
\be
  \cL_{\rm bd} = \brc{e^{-2\sgm(\pi R)}\int d^4\tht\; \abs{\chi}^2
 +e^{-3\sgm(\pi R)}\sbk{\int d^2\tht\;
 \brkt{\xi H^c\chi+\frac{1}{2}m_\chi\chi^2}+\hc}}\dlt(y-\pi R). 
\ee
where $\xi$ and $m_\chi$ are complex parameters 
 whose mass-dimensions are $1/2$ and $1$. 
Through the 
 similar calculations to the Case~4, we obtain 
 an equation that determines the mass spectrum. 
It is obtained from
 Eq.(\ref{det_cM2}) by the replacement,  
\be
 (\tl{S}_{\pm M_H},\tl{C}_{\pm M_H}) \to 
 (\tl{C}_{\mp M_H},-\tl{S}_{\mp M_H}). 
\ee
Then the mass spectrum is determined by  
\be
 \tl{T}_{M_H}^{-1} = 
 -\frac{\abs{\xi}^2}{2\brkt{e^{\sgm(\pi R)}m_n\pm\abs{m_\chi}}}. 
\ee

\section{Bases of mode functions} \label{bases_fcn}

This section shows
 the definitions and properties 
 of functions, $C(y,m)$, $S(y,m)$, $C_M(y,m)$, 
 and $S_M(y,m)$, following Ref.\cite{Falkowski}. 
The functions~$C(y,m)$ and $S(y,m)$
 are defined as solutions of 
\be
 \brc{\der_y^2-2\sgm'\der_y+m^2e^{2\sgm}}f = 0 
 \label{def_eq:gauge}
\ee
with initial conditions of  
\bea
 C(0,m) \eql 1, \;\;\;\;\;
 C'(0,m) = 0, \nonumber\\
 S(0,m) \eql 0, \;\;\;\;\;
 S'(0,m) = m. 
\eea
From the Wronskian relation, they satisfy 
\be
 S'(y,m)C(y,m)-C'(y,m)S(y,m) = me^{2\sgm(y)}. 
\ee

Next we provide the definition of $C_M(y,m)$ and $S_M(y,m)$. 
Combining the two equations in Eq.(\ref{bulk:md_eq1}), we obtain 
 the following type of the second order differential equation, 
\be
 \brc{\der_y^2-\sgm'\der_y-M(M+\sgm')+m^2e^{2\sgm}}f_M = 0. 
 \label{def_eq1}
\ee
By redefining $f_M(y)$ as 
\be
 \tl{f}_M(y) \equiv e^{My}f_M(y),  
\label{def_hatf}
\ee
Eq.(\ref{def_eq1}) becomes 
\be
 \brc{\der_y^2-(\sgm'+2M)\der_y+m^2e^{2\sgm}}\tl{f}_M = 0. 
 \label{def_eq2}
\ee
The functions~$C_M(y,m)$ and $S_M(y,m)$ are 
 solutions of Eq.(\ref{def_eq2}) which satisfy 
 initial conditions,         
\bea
 C_M(0,m) \eql 1, \;\;\;\;\;
 C_M'(0,m) = 0, \nonumber\\ 
 S_M(0,m) \eql 0, \;\;\;\;\;
 S_M'(0,m) = m.  
 \label{ini_cond}
\eea

Here let us define a function~$\tl{g}_M(y)$ as 
\be
 \tl{g}_M(y) \equiv e^{-\sgm(y)-2My}\tl{f}'_M(y). 
\ee
Then it satisfies  
\be
 \brc{\der_y^2-(\sgm'-2M)\der_y+m^2e^{2\sgm}}\tl{g}_M = 0. 
\ee
This means $\tl{g}_M(y)\propto\tl{f}_{-M}(y)$. 
Taking into account the initial conditions, we obtain 
\bea
 C'_M(y,m) \eql -me^{\sgm+2My}S_{-M}(y,m), \nonumber\\
 S'_M(y,m) \eql me^{\sgm+2My}C_{-M}(y,m). 
 \label{der_fml}
\eea
Furthermore, using the Wronskian relation, we obtain  
\be
 S_M'(y,m)C_M(y,m)-C_M'(y,m)S_M(y,m) = me^{\sgm(y)+2My}, 
\ee
which is translated into 
\be
 C_M(y,m)C_{-M}(y,m)+S_M(y,m)S_{-M}(y,m) = 1,
  \label{CS_rel1}
\ee
by using Eq.(\ref{der_fml}).

\subsection{Flat spacetime}

Let us see explicit forms of $C_M(y,m)$ and $S_M(y,m)$ 
 in the case of the flat spacetime, \ie, $\sgm(y)=0$. 
In this case, 
 Eq.(\ref{def_eq:gauge}) is reduced to 
\be
 \brkt{\der_y^2+m^2}f=0, 
\ee
and 
\be
 C(y,m) = \cos(my), \;\;\;\;\;
 S(y,m) = \sin(my). 
\ee
On the other hand, 
 Eq.(\ref{def_eq1}) becomes 
\be
 \brkt{\der_y^2-M^2+m^2}f_M = 0. 
\ee
Thus a general solution is 
 given by
\be
 f_M(y) = A\cos(\sqrt{m^2-M^2}y)+B\sin(\sqrt{m^2-M^2}y), 
\ee
where $A$ and $B$ are integration constants, and  
 $m^2\geq M^2$ is assumed. 
{}From Eqs.(\ref{def_hatf}) and (\ref{ini_cond}), 
 we obtain 
\bea
 C_M(y,m) \eql e^{My}\brc{\cos(\sqrt{m^2-M^2}y)
 -\frac{M}{\sqrt{m^2-M^2}}\sin(\sqrt{m^2-M^2}y)} \nonumber\\
 \eql \frac{m}{\sqrt{m^2-M^2}}e^{My}\cos(\sqrt{m^2-M^2}y+\vph), \nonumber\\
 S_M(y,m) \eql \frac{m}{\sqrt{m^2-M^2}}e^{My}\sin(\sqrt{m^2-M^2}y), 
\eea
where 
\be
 \vph \equiv \arctan\brkt{\frac{M}{\sqrt{m^2-M^2}}}. 
\ee

\subsection{Randall-Sundrum spacetime}

Finally we consider the case of the warped background setup
 of $\sgm(y)=ky$ ($k>0$, $0\leq y\leq \pi R$). 
In this case solutions of Eq.(\ref{def_eq2}) are expressed by 
 the Bessel functions as 
\bea
 C_M(y,m) \eql \frac{\pi m}{2k}e^{\alp ky}
 \brc{Y_{\alp-1}\brkt{\frac{m}{k}}J_\alp\brkt{\frac{m}{k}e^{ky}}
 -J_{\alp-1}\brkt{\frac{m}{k}}Y_\alp\brkt{\frac{m}{k}e^{ky}}} \nonumber\\
 \eql \frac{\pi m}{2k}\frac{e^{\alp ky}}{\sin\pi\alp}
 \brkt{J_{1-\alp}\brkt{\frac{m}{k}}J_\alp\brkt{\frac{m}{k}e^{ky}}
 +J_{\alp-1}\brkt{\frac{m}{k}}J_{-\alp}\brkt{\frac{m}{k}e^{ky}}}, \nonumber\\
 S_M(y,m) \eql -\frac{\pi m}{2k}e^{\alp ky}
 \brc{Y_\alp\brkt{\frac{m}{k}}J_\alp\brkt{\frac{k}{m}e^{ky}}
 -J_\alp\brkt{\frac{m}{k}}Y_\alp\brkt{\frac{m}{k}e^{ky}}} \nonumber\\
 \eql \frac{\pi m}{2k}\frac{e^{\alp ky}}{\sin\pi\alp}
 \brc{J_{-\alp}\brkt{\frac{m}{k}}J_\alp\brkt{\frac{m}{k}e^{ky}}
 -J_\alp\brkt{\frac{m}{k}}J_{-\alp}\brkt{\frac{m}{k}e^{ky}}}, 
\eea
where $\alp\equiv (M/k)+1/2$. 
We can see 
\be
 C(y,m) = C_{k/2}(y,m), \;\;\;\;\;
 S(y,m) = S_{k/2}(y,m), 
\ee
{}from Eqs.(\ref{def_eq:gauge}) and (\ref{def_eq2}).


\end{document}